\documentclass[12pt]{article}
\usepackage{epsfig}
\usepackage{graphicx}
\usepackage{cite}
\usepackage{amsfonts}
\usepackage{amssymb}
\usepackage{bm}
\usepackage{latexsym}
\usepackage{amsmath}
\numberwithin{equation}{section}
\def\ee{\end{equation}}
\def\ba{\begin{eqnarray}}
\def\ea{\end{eqnarray}}

\def\bq{\begin{quote}}
\def\eq{\end{quote}}

 at 10truept

\newcommand{\beq}{\begin{equation}}
\newcommand{\eeq}{\end{equation}}
\newcommand{\beqa}{\begin{eqnarray}}
\newcommand{\eeqa}{\end{eqnarray}}
\newcommand{\bea}{\begin{eqnarray}}
\newcommand{\eea}{\end{eqnarray}}

 \newcommand{\be}{\beta}
 \newcommand{\ep}{\epsilon}

\newcommand{\vect}[1]{\bm{\mathrm{{#1}}}}

\def\baq{\begin{eqnarray}}
\def\eaq{\end{eqnarray}}

\def\fnl{f_{\rm NL}}

\def\q{{\bf q}}
\def\x{{\bf x}}
\def\k{{\bf k}}
\def\d{{\rm d}}

\def\B{{\rm B}}
\def\bnl{b_{\rm NL}}
\def\cnl{c_{\rm NL}}

\def\lesssim{~\mbox{\raisebox{-.6ex}{$\stackrel{<}{\sim}$}}~}

\def\ltap{\ \raise.3ex\hbox{$<$\kern-.75em\lower1ex\hbox{$\sim$}}\ }
\def\gtap{\ \raise.3ex\hbox{$>$\kern-.75em\lower1ex\hbox{$\sim$}}\ }
\def\gl{\ \raise.5ex\hbox{$>$}\kern-.8em\lower.5ex\hbox{$<$}\ }
\def\roughly#1{\raise.3ex\hbox{$#1$\kern-.75em\lower1ex\hbox{$\sim$}}}

\begin{document}

\begin{titlepage}

\nopagebreak

\title{\bf Constraints on Gauge Field Production\\ during Inflation}

\vfill
\author{Sami Nurmi$^{a}$\footnote{sami.nurmi@helsinki.fi}\ \  and Martin S. Sloth$^{b}$\footnote{sloth@cp3.dias.sdu.dk}
}
\date{ }


\maketitle

\vskip 0.5cm

\begin{center}
{\it  $^{a}$  University of Helsinki and Helsinki Institute of
Physics,}\\
{\it P.O. Box 64, FI-00014, Helsinki, Finland}

\vskip 0.5cm
{\it  $^{b}$ CP$^3$-Origins, Centre for Cosmology and Particle Physics Phenomenology,} \\
{\it  University of Southern Denmark, Campusvej 55, 5230 Odense M, Denmark}

\thispagestyle{empty}
\end{center}
\vfill
\begin{abstract}

In order to gain new insights into the gauge field couplings in the early universe, we consider the constraints on gauge field production during inflation imposed by requiring that their effect on the CMB anisotropies are subdominant. In particular,
we calculate systematically the bispectrum of the primordial curvature perturbation induced by the presence of vector gauge fields during inflation. Using a model independent parametrization in terms of magnetic non-linearity parameters, we calculate for the first time the contribution to the bispectrum from the cross correlation between the inflaton and the magnetic field defined by the gauge field. We then demonstrate that in a very general class of models, the bispectrum induced by the cross correlation between the inflaton and the magnetic field can be dominating compared with the non-Gaussianity induced by magnetic fields when the cross correlation between the magnetic field and the inflaton is ignored.

 \end{abstract}
 \vskip.4in

\hfill \\
\vfill
\end{titlepage}

\setcounter{equation}{0} \setcounter{footnote}{0}

\section{Introduction}

Recent data from the Planck satellite has verified the paradigm of single field slow roll inflation to unprecedented high precision \cite{Ade:2013uln,Ade:2013ydc}. This alone is a great success, but it also provides new nontrivial constraints on other degrees of freedom, which we either know are there in the post-inflationary universe (neutrinos \cite{Lesgourgues:2006nd,Hannestad:2010kz,Wong:2011ip,Lesgourgues:2012uu,Abazajian:2011dt}, magnetic fields \cite{Grasso:2000wj,Giovannini:2003yn,Kandus:2010nw,Yamazaki:2012pg},
dark matter \cite{Archidiacono:2013cha,Hutsi:2011vx,Galli:2011rz,Delahaye:2011jn,Hamann:2006pf}, dark energy \cite{Jassal:2006gf,Sherwin:2011gv,Ade:2013dsi,Dodelson:2013pln}, etc.) or which we have some reasons to believe could be present even during inflation, such as in multi-field models of inflation \cite{Langlois:2008mn,Byrnes:2010em}, the curvaton model \cite{Enqvist:2001zp,Lyth:2001nq,Moroi:2001ct}, or models of inflationary magnetogenesis \cite{Turner:1987bw,Ratra:1991bn,Dolgov:1993vg,Davis:2000zp,Giovannini:2002ki,Bertolami:2005np,Bamba:2006ga,Kanno:2009ei}.  In the present paper we will focus on the constraints imposed on a $U(1)$ vector field coupled to the inflaton, coming from the observational constraints on non-Gaussianity in the CMB. For the applications in this paper, this could be a completely general $U(1)$ vector field, but the applications of the results presented below are especially interesting when one identifies the vector field with the one of elect!
 romagnetism.

Cosmic magnetic fields with a coherent scale as large as 100 kpc and a strength of order $\mu$G has been established to be present in galaxies and clusters of galaxies \cite{Kronberg:1993vk,Clarke:2000bz,Carilli:2001hj}. It is believed that the origin of such magnetic fields might be due to an enhancement of pre-existing small magnetic fields, called seed fields, due to the dynamo mechanism. It is generally assumed that these seed fields need to have a strength larger than about $10^{-20}$ Gauss in order for the dynamo mechanism to work \cite{Turner:1987bw}, although it has been claimed that this lower bound can be significantly relaxed in the presence of dark energy \cite{Davis:1999bt}. Two possible explanations for the origin of such seeds exists. One is the possibility that conformal invariance of electro-magnetism is broken sufficiently during inflation, in order to enhance quantum fluctuations of the $U(1)$ vector field and generate the seed magnetic fields at the end o!
 f inflation \cite{Turner:1987bw,Ratra:1991bn,Dolgov:1993vg,Davis:2000zp,Giovannini:2002ki,Bertolami:2005np,Bamba:2006ga}. Another possibility  is that the seeds are generated after inflation, f.ex. during a phase transition or from the Bierman battery mechanism \cite{Kandus:2010nw}. Recently there has however been a claimed indirect observation of femto Gauss magnetic fields with a coherence length of super Mpc scales \cite{Neronov:1900zz,Taylor:2011bn,Vovk:2011aa}. If this is true, this could pose a problem for mechanisms which generates the magnetic seeds by causal processes after inflation, since the coherence length of such magnetic fields are limited by the horizon at the time of generation, and is typically too small to explain magnetic fields with a coherence length larger than the Mpc scale. This might suggest that the magnetic seeds are generated during inflation.

However, it is a challenge for inflationary magnetogenesis to
identify a source that breaks sufficiently the conformal invariance
of electro-magnetism during inflation. If the conformal invariance
would be unbroken, then the vector field perturbation would not get
enhanced during inflation, and no significant magnetic fields would
be generated. One of the simplest and most popular models  for
breaking the conformal invariance during inflation is to add a
non-minimal coupling between the gauge kinetic term
$F_{\mu\nu}F^{\mu\nu}$ and the inflaton, $\phi$, of the form
$\lambda(\phi)F_{\mu\nu}F^{\mu\nu}$
\cite{Turner:1987bw,Ratra:1991bn,Dolgov:1993vg,Davis:2000zp,Giovannini:2002ki,Bertolami:2005np,Bamba:2006ga}.
Backreaction provides a simple constraint on the magnetic field
generated during inflation in this type of models, although in
principle inflationary attractor solutions dominated by the gauge
field energy density may also exist \cite{Dimopoulos:2010xq}. In
order not to disturb the standard inflationary picture we must
require that the energy density in the magnetic field $\rho_B$ is
smaller than the total energy density $\rho$ during inflation, while
at the same time staying in a perturbative regime in order to avoid
the strong coupling problem \cite{Demozzi:2009fu,Ferreira:2013sqa}.
In fact, it has been demonstrated that the generation of significant
seed magnetic fields from inflation seems to require low-scale
inflation \cite{Ferreira:2013sqa}. However, the fluctuating magnetic
field also contributes to the total curvature perturbation, $\zeta$,
and since the perturbations from the magnetic field are non-Gaussian
this leads to additional strong constraints on the strength of
magnetic fields generated during inflation \cite{Barnaby:2012tk,
Lyth:2013sha,Fujita:2013qxa}.

In addition, due to the non-minimal coupling between the inflaton and the vector field, models of the type  $\lambda(\phi)F_{\mu\nu}F^{\mu\nu}$ will also induce non-trivial correlations between the inflation fluctuations and the magnetic field. Such cross correlations was recently studied in \cite{Caldwell:2011ra,Motta:2012rn,Jain:2012ga,Jain:2012vm,Biagetti:2013qqa,Shiraishi:2012xt}, and in \cite{Jain:2012ga} it was suggested that such cross correlations could be parametrized in a model independent way in terms of a magnetic non-linearity parameter, $b_{NL}$ of the form $\left< \zeta~{\bf B}\cdot {\bf B}\right> \propto  b_{NL} P_{\zeta} P_{B}$ analogous to the definition of $f_{NL}$, where here $P_{\zeta}$ and $P_{B}$ are the power spectra of the curvature perturbation and the magnetic field respectively. In fact one can derive a new ``magnetic consistency relation'' in terms of the parameter $b_{NL}$ \cite{Jain:2012ga,Jain:2012vm}

In the work presented here, we will analyze the induced
non-Gaussianity in the CMB from such cross-correlations between the
inflation fluctuations and the magnetic field in the general class
of models where the gauge field action takes the form
  \beq
  \label{Lgauge}
  {\cal L}_{\rm gauge}=\lambda(\phi)F_{\mu\nu}F^{\mu\nu}\ .
  \eeq
The general analysis allows to use the level of induced
non-Gaussianity to constrain the possible forms of the coupling
$\lambda(\phi)$. As a benchmark model, we will also calculate the
induced non-Gaussianity from cross-correlations in the extensively
studied models where the coupling $\lambda(\phi)$ takes a power law
form. This new contribution which will turn out to be the dominant
non-Gaussian contribution in certain shapes.

As already mentioned above, the non-Gaussianity induced by the
magnetic field, when ignoring the cross-correlations with the
inflaton, has already been studied extensively the literature in the
specific $\lambda(\phi)F_{\mu\nu}F^{\mu\nu}$ models with a power law
coupling $\lambda(\phi)$ \cite{Barnaby:2012tk,
Lyth:2013sha,Fujita:2013qxa}.  In order to understand the relation
between the different results in the literature and the results
presented here, let us write the total curvature perturbation in
terms of the curvature perturbation in the inflaton fluid,
$\zeta_\phi$, and in the magnetic field fluid, $\tilde\zeta_B$,
as\footnote{Note that our $\tilde\zeta_B$ is ignored in the analysis
of both \cite{Barnaby:2012tk, Lyth:2013sha}, where only the sourcing
of $\zeta_{\phi}$ from the interaction of $\phi$ with the vector
field is considered, which is an inconsistent approximation as we
will now see.}
\beq
\label{ztotal}
\zeta = -H \frac{\delta\rho_\phi +\delta\rho_B}{\dot\rho} \equiv
\zeta_\phi + \tilde\zeta_B\ .
\eeq
At the background level $\rho=\rho_{\phi}$ as we assume a vanishing
v.e.v. for the magnetic field. However, at first order in
perturbations the average energy density of the magnetic field
fluctuations gives an effective background component
$\rho_{B}\equiv\langle\delta\rho_B\rangle$. Considering fluctuations
over the average value we can define the intrinsic curvature
perturbation of the magnetic fluid as
  \beq
  \hat{\zeta}_{B}= \frac{\dot{\rho}_{\phi}}{\dot{\rho}_B}\tilde{\zeta}_{B}\ .
  \eeq

Consider the time derivative of $\zeta$, to see how it grows with
time. It is well known that in the absence of direct coupling
between the fluids, the curvature perturbation in each fluid is
separately conserved on superhorizon scales and we have
$\dot{\zeta}_\phi\simeq\dot{\hat\zeta}_B\simeq 0$, while in the
presence of sources we have \bea\label{Q-Q}
\dot{\delta\rho_\phi} +3H(\delta\rho_\phi+\delta p_\phi) &=&-Q\nonumber\\
\dot{\delta\rho_B} +3H(\delta\rho_B+ \delta p_B) &=&Q\ . \eea From
the continuity equation of the electromagnetic field in the regime
where the magnetic fields dominates the electromagnetic energy
density, we have \cite{Jain:2012vm}
\beq
\dot{\delta\rho_B}+4H\delta\rho_B=\frac{\dot{\lambda}}{\lambda}\delta\rho_B\
,
\eeq
it follows that the energy transfer term is given by
\beq
Q=\frac{\dot{\lambda}}{\lambda}\delta\rho_B\ .
\eeq

From (\ref{Q-Q}) it follows that
\bea
\label{dotzphi}
\dot{\zeta}_\phi &=& 3\frac{H^2}{\dot\rho_\phi}(\delta p_\phi
-\frac{\dot p_\phi}{\dot\rho_\phi}\delta\rho_\phi)
+H\frac{Q}{\dot\rho_\phi}\approx H\frac{Q}{\dot{\rho}_{\phi}}\nonumber \\
\dot{\hat\zeta}_B &=& 3\frac{H^2}{\dot\rho_B}(\delta p_B -\frac{\dot
p_B}{\dot\rho_B}\delta\rho_B)-H\frac{Q}{\dot\rho_B}- \frac{\dot
Q}{\dot\rho_B}\hat\zeta_B \approx -H\frac{Q}{\dot\rho_B} -
\frac{\dot Q}{\dot\rho_B}\hat\zeta_B \
  \eea
where in the last steps we assumed that the intrinsic non-adiabatic
pressure in the two fluids vanishes. We have also neglected slow
roll suppressed terms proportional to ${\dot H}$.

Now we can compute $\dot {\tilde\zeta}_B$, which gives
\bea
\label{dotzb}
\dot{\tilde\zeta}_B
&=& -H\frac{Q}{\dot{\rho}_{\phi}} -\frac{H}{\rho_{\phi}+p_{\phi}}\left( \delta p_B
-\frac{\dot{p}_{\phi}}{\dot{\rho}_{\phi}}\delta\rho_B\right)\nonumber\\
&\equiv&  -H\frac{Q}{\dot{\rho}_{\phi}}
-\frac{H}{{\rho}_{\phi}+{p}_{\phi}}\delta P_{nad}~. \eea Thus,
clearly if we were to compute $\dot\zeta$, the source term, $Q$,
cancels out and we obtain
 \beq\label{zdot}
\dot\zeta =\dot\zeta_\phi + \dot{\tilde\zeta}_B =
-\frac{H}{\rho+p}\delta P_{nad}.
\eeq
This is in agreement with \cite{Fujita:2013qxa}, but in general
inconsistent with assuming $ \dot{\tilde\zeta}_B=0$ and considering
only the source term on $\zeta_\phi$ as in  \cite{Barnaby:2012tk,
Lyth:2013sha} (see appendix \ref{source} for further discussion of
this point). From equation (\ref{dotzphi}) we see that the curvature
perturbations of the inflaton fluid and the magnetic fluid evolve
only if the coupling $\lambda$ is changing in time. However, as the
total curvature perturbation (\ref{ztotal}) is not just a sum of
$\zeta_{\phi}$ and $\hat\zeta_{B}$ but their sum weighted by the
ratios of the individual fluid energies the curvature perturbation
evolves even if $\zeta_{\phi}$ and $\hat\zeta_{B}$ are constant but
the fluid energies $\rho_{\phi}$ and $\rho_{B}$ evolve differently.

As we will discuss in section 4, the non-adiabatic pressure, $\delta
P_{nad}$, is proportional to the strength of the magnetic field
squared, $B^2$. By integration of equation (\ref{zdot}), we see that
there are two distinct contributions to the curvature perturbation.
There is a contribution proportional to the magnetic field squared,
$B^2$, obtained by integrating the the non-adiabtic pressure on
super-horizon scales, which we will label $\zeta_B$. In addition
there is a constant of integration which is the contribution to the
curvature perturbation at horizon crossing, which is independent of
the magnetic field and given by the inflation fluctuation. We will
label this constant of integration $\zeta_0$. We can the write the
total curvature perturbation simply as
\beq
\label{zetasplitintro}
\zeta =\zeta_0 + \zeta_B~,
\eeq
where $\zeta_0$ is given by $\zeta_{\phi}$ evaluated at horizon crossing, and $\zeta_B$ is the super-horizon contribution determined by the non-adiabatic pressure, which is proportional to $B^2$.

While the correlation function
\beq
\left<\zeta_B \zeta_B \zeta_B\right>
\eeq
which contributes to the observable $\langle\zeta\zeta\zeta\rangle$,
parameterized by the non-linearity parameter $f_{NL}$, was computed
in \cite{Fujita:2013qxa} (see also \cite{Barnaby:2012tk,
Lyth:2013sha}), the correlation between $\zeta_0$ and $\zeta_B$ is
to our knowledge neglected in all of the previous work. As can be
seen from equation (\ref{zetasplitintro}), the three point function
of the total curvature perturbation $\zeta$ also receives
contributions from terms of the form
\beq\label{cor1}
\left<\zeta_0 \zeta_B \zeta_B\right> ~, \qquad \left<\zeta_0 \zeta_0
\zeta_B\right>~.
\eeq
The main point of this paper is to calculate the cross correlation
contributions. In section 4, we will see that these terms can give
the dominant contribution to the observable
$\langle\zeta\zeta\zeta\rangle$, even larger than the contribution

from $\left<\zeta_B \zeta_B \zeta_B\right>$ computed in
\cite{Fujita:2013qxa}. Since $\delta P_{nad}$ is proportional to the
strength of the magnetic field squared, $B^2$, the correlators of
the type shown in (\ref{cor1}) will be given in terms of cross
correlation function of the magnetic field with the curvature
perturbation
\beq
\left<\zeta_0 \zeta_0 B^2\right> ~, \qquad \left<\zeta_0  B^2 B^2 \right>~.
\eeq
In a specific model these correlators will have to be computed in the {\it in-in} formalism \cite{Giddings:2010ui} going beyond linear perturbation theory, which for every new model can a tedious calculation. However, in the next section, we will discuss how these correlation functions can be parametrized in terms of magnetic non-linearity parameters in a model independent way, and in section 3 we will show how to evaluate them.

The paper is organized as follows. In the next section introduce the
magnetic non-linearity parameters, and show how the cross
correlation functions of the curvature perturbation with the
magnetic field can be parametrized in a model independent way. In
section 3 we evaluate these model independent cross-correlation
functions. In section 4, we find the induced non-Gaussinity from the
cross correlation functions, and in section 5 we also consider the
size of these cross correlation functions in the specific models
where the coupling $\lambda(\phi)$ in
$\lambda(\phi)F_{\mu\nu}F^{\mu\nu}$ takes a power law form. Finally,
in section 6, we conclude and summarize our results.

\section{The magnetic non-linearity parameters}

If we define the cross-correlation bispectrum of the curvature perturbation with the magnetic fields as
\beq \label{zBB0}
\left<\zeta_0(\vect{k}_1){\bf B}(\vect{k}_2)\cdot {\bf B}(\vect{k}_3)\right> \equiv(2\pi)^3\delta^{(3)}(\vect{k}_1+\vect{k}_2+\vect{k}_3)B_{\zeta B B}(\vect{k}_1,\vect{k}_2, \vect{k}_3) ~,
\eeq
then is has previously been proposed, that it is convenient to define the magnetic non-linearity parameter $b_{NL}$, in terms of the cross-correlation function of the curvature perturbation with the magnetic fields
\beq \label{zBB1}
B_{\zeta B B}(\vect{k}_1,\vect{k}_2, \vect{k}_3) \equiv \frac{1}{2}b_{NL}P_\zeta(k_1)\left( P_B(k_2)+P_B(k_3)\right)~,
\eeq
where $P_\zeta$ and $P_B$ are the power spectra of the comoving curvature perturbation and the magnetic fields,
defined respectively as
\bea
\left<\zeta(\vect{k})\zeta(\vect{k}')\right> \equiv(2\pi)^3\delta^{(3)}(\vect{k}+\vect{k}') P_\zeta(k),\\
\left<{\bf B}(\vect{k})\cdot {\bf B}(\vect{k}')\right> \equiv(2\pi)^3\delta^{(3)}(\vect{k}+\vect{k}') P_B(k).
\eea

Similarly we may also introduce the magnetic trispectrum
\beq \label{zzBB0}
\left<\zeta_0(\vect{k}_1)\zeta_0(\vect{k}_2){\bf B}(\vect{k}_3)\cdot {\bf B}(\vect{k}_4)\right> \equiv(2\pi)^3\delta^{(3)}(\vect{k}_1+\vect{k}_2+\vect{k}_3+\vect{k}_4)T_{\zeta\zeta B B}(\vect{k}_1,\vect{k}_2, \vect{k}_3,\vect{k}_4) ~,
\eeq
which can be parametrized in terms of new magnetic non-linearity parameters $\be_{NL}$ and $c_{NL}$,
\beq \label{zzBB1}
T_{\zeta\zeta B B}(\vect{k}_1,\vect{k}_2,\vect{k}_3 ,\vect{k}_4) \equiv\be_{NL} P_\zeta(k_1) P_\zeta(k_2) P_B(k_{13})+\frac{2}{3}c_{NL}P_\zeta(k_1) P_\zeta(k_2) P_B(k_4)~+perm.,
\eeq

In the case where $b_{NL}$ is momentum independent and quantum interference effects around horizon crossing can be ignored, it takes a ``local" form which can be derived from the relation
\beq
\label{Blocal} {\bf B} = {\bf B}^{(G)} +\frac{1}{2}b_{NL}^{local} \zeta_0
{\bf B}^{(G)}+\frac{1}{6}c_{NL}^{local} \zeta_0  \zeta_0
{\bf B}^{(G)}
\eeq
with ${\bf B}^{(G)}$ and $\zeta_0$ being Gaussian fields. With this local ansatz one obtains that the $\be_{NL}$ term in the trispectrum is given by
\beq
\be^{local}_{NL} =  \frac{1}{2}(b^{local}_{NL})^2~.
\eeq

There are interesting limits where indeed the magnetic bispetrum and trispectrum can be derived from semiclassical considerations, and in these "squeezed" limits the magnetic non-linearity parameter takes the local form. It has previously been shown that in the squeezed limit, where the  momentum of the curvature perturbation vanishes, i.e., $k_1\ll k_2, k_3$, the bispectrum in fact takes the form
\beq
\left<\zeta_0(\vect{k}_1){\bf B}(\vect{k}_2)\cdot {\bf B}(\vect{k}_3)\right> =b_{NL}^{local}(2\pi)^3\delta^{(3)}(\vect{k}_1+\vect{k}_2+\vect{k}_3)P_\zeta(k_1) P_B(k_2)~,
\eeq
with $ b_{NL}^{local} = n_B-4$ where $n_B$ is the spectral index of the magnetic field power spectrum, in agreement with the magnetic consistency relation, which was derived in \cite{Jain:2012ga,Jain:2012vm}. In the case of a scale invariant spectrum of magnetic fields, $n_B= 0$, we have $b^{local}_{NL} = -4$ (see also appendix \ref{appPB}).

Another interesting limit which maximizes the three-point cross-correlation function is the flattened shape where $k_1/2=k_2=k_3$. In this limit it turns out that the signal is enhanced by a logarithmic factor in agreement with \cite{Caldwell:2011ra,Motta:2012rn,Jain:2012ga,Jain:2012vm}. On the largest scales the logarithm will give an enhancement by a factor 60. Thus, for a flat magnetic field power spectrum, the non-linearity parameter in the flattened limit becomes $|b_{NL}|  \sim  \mathcal O(10^3) $ depending on the scale.

\section{Three-point cross-correlation functions}

\label{sec:zbcorrelators}

Since the electromagnetic part of the perturbed action contains only
terms of the form $A^2\zeta^n$, see equation (\ref{Lgauge}), the
curvature perturbation generated the magnetic fields is of the form
$\zeta_{B}\propto B_{i}B^{i}/(H^2M_{\rm P}^2)$. The magnetic fields
generated during inflation obey a Gaussian statistics to leading
order in perturbations so that the induced curvature perturbation
$\zeta_{\B}$ is a non-Gaussian field.

To estimate the contribution of magnetic fields to the bispectrum of
primordial density fluctuations we should consider three-point
functions of the form.
  \beq
  \langle\zeta_0(\vect{k}_1)\zeta_0(\vect{k}_2) {\bf B}^2(\vect{k}_3)\rangle\ ,\qquad \langle\zeta_0(\vect{k}_1) {\bf B}^2(\vect{k}_2)
  {\bf B}^2(\vect{k}_3)\rangle\ ,\qquad \langle {\bf B}^2(\vect{k}_1) {\bf B}^2(\vect{k}_2)
  {\bf B}^2(\vect{k}_3)\rangle\ .
  \eeq
To lowest order in perturbations, the amplitudes of the two first
correlators depend on the parameters $\bnl$ and $\cnl$ in the
expansion (\ref{Blocal}) while the last correlator only depends on
the amplitude of magnetic fields.

The two-point function of the magnetic fields is given to lowest
order in perturbations by,
  \beq
  \langle
  B_{i}(\k)B_{j}(\k')\rangle=(2\pi)^3\delta(\k+\k')\frac{1}{2}\left(\delta_{ij}-\frac{k_i
  k_j}{k^2}\right)P_{\B}(k)\ .
  \eeq

For simplicity, we assume the scale and time dependence of the
magnetic spectrum can be parameterized by a power law as
  \beq
  \label{P_B1}
  P_{B}(k)=\frac{C_B}{\lambda(\eta)} (-k\eta)^{4-2n}k^{-3}\ ,
  \eeq
where $C_B$ is a constant. The power law spectrum of magnetic fields
is obtained in the extensively studied class of models with ${\cal
L} = \lambda(\phi)F_{\mu\nu}F^{\mu\nu}$ and a power law form for the
coupling $\lambda\propto a^{n}$. However, in this case the
coefficients $\bnl$ and $\cnl$, which are determined by the
derivatives of the coupling $\lambda(\phi)$ \cite{Jain:2012ga}, are
not independent free parameters, and therefore constraints on
$\bnl$ and $\cnl$ are of limited use. In the general case where
$\bnl$ and $\cnl$ are treated as free parameters, but the form of
the spectrum is still assumed to take the power-law form
(\ref{P_B1}), it is then evident, that we are implicitly
concentrating on a limited class of models. However, it can be shown that an
approximatively power law spectrum can be obtained in models with
${\cal L} = \lambda(\phi)F_{\mu\nu}F^{\mu\nu}$ for couplings of the
form $\lambda(a)=\lambda(a_0)(1-\bnl {\rm ln}(a/a_0)+\ldots)$ (see
appendix \ref{appPB}). More generally one could think that for
example deviations from Bunch-Davies vacuum or models with extra
degrees of freedom could effectively yield a power law spectrum for
magnetic fields while still featuring the coefficients $\bnl$ and
$\cnl$ in (\ref{Blocal}) as independent parameters.

With this being said we will here adopt a purely phenomenological
approach simply assuming the magnetic spectrum takes a power law
form and investigating the constraints on the $\bnl$ and $\cnl$ in
the parametrization (\ref{Blocal}). The case $n=2$ in (\ref{P_B1})
then corresponds to a scale-invariant spectrum $P_{B}\propto
k^{-3}$. Here we will concentrate on the regime $n>-1/2$ to
eventually connect with the regime of strongly coupled magnetic
fields. The results in the other regime can be obtained by use of
the electromagnetic duality, which with the power-law assumption for $\lambda$ leaves the result invariant under
a simultaneous exchange of the electric and magnetic field and $n\to
-n$ \cite{Buonanno:1997zk,Brustein:1998kq,Giovannini:2009xa}.

\subsection{Correlators of the form $\langle\zeta_0\zeta_0 B^2\rangle$}

Using the definition (\ref{zzBB1}), we find to lowest order in
perturbations the result
  \beq
  \langle\zeta_0(\k_1)\zeta_0(\k_2){\bf B}^2(\k_3)\rangle_{\rm c} =
  (2\pi)^3\delta(\k_1+\k_2+\k_3)P_{\zeta}(k_1)P_{\zeta}(k_2)\left(\beta_{NL}+\frac{2}{3}\cnl\right)
  \hspace{-20pt}\int\limits_{k_0<q<aH}\hspace{-20pt}\frac{\d\q}{(2\pi)^3}P_{\B}(q)+2{\rm p.}
  \eeq
which with the local ansatz becomes
  \beq
  \label{zzbb_int}
  \langle\zeta_0(\k_1)\zeta_0(\k_2){\bf B}^2(\k_3)\rangle_{\rm c} =
  (2\pi)^3\delta(\k_1+\k_2+\k_3)P_{\zeta}(k_1)P_{\zeta}(k_2)\left(\frac{1}{2}(\bnl^{local})^2+\frac{2}{3}\cnl^{local}\right)
  \hspace{-20pt}\int\limits_{k_0<q<aH}\hspace{-20pt}\frac{\d\q}{(2\pi)^3}P_{\B}(q)+2{\rm p.}
  \eeq
Here $k_0=a_0H_0$ denotes the horizon scale at the onset of magnetic
field generation and $aH$ is the horizon scale at the time when the
correlators are evaluated. We have only included the connected part
of the correlator and $P_{\zeta}$ denotes the spectrum of the
Gaussian part of curvature perturbations generated independently of
the magnetic fields.

Using the expression (\ref{P_B1}) for the magnetic spectrum, the
integral in (\ref{zzbb_int}) can be easily computed and one finds
  \baq
  \label{zzbb_sol}
  \langle\zeta_0(\k_1)\zeta_0(\k_2){\bf B}^2(\k_3)\rangle_{\rm c}
  &=&
  (2\pi)^3\delta(\k_1+\k_2+\k_3)P_{\zeta}(k_1)P_{\zeta}(k_2){\cal P}_{B}(k_0)
  \times
  \\\nonumber&&\left(\frac{1}{2}(\bnl^{local})^2+\frac{2}{3}\cnl^{local}\right)
  \frac{1}{4-2n}\left(\left(\frac{aH}{k_0}\right)^{4-2n}-1\right)+2{\rm
  p.}
  \eaq
In the scale-invariant case $n=2$ this reduces to the form
 \baq\label{zzb2}
  \langle\zeta_0(\k_1)\zeta_0(\k_2){\bf B}^2(\k_3)\rangle_{\rm c}
  &=&
  (2\pi)^3\delta(\k_1+\k_2+\k_3)P_{\zeta}(k_1)P_{\zeta}(k_2)\mathcal{P}_{\B}(k_0)
   \times
  \\\nonumber&&
  \left(\frac{1}{2}(\bnl^{local})^2+\frac{2}{3}\cnl^{local}\right) \ln\left(\frac{aH}{k_0}\right)+2{\rm
  p.}
  \eaq

\subsection{Correlators of the form $\langle\zeta_0 B^2 B^2\rangle$}

As the Lagrangian does not contain higher order terms in the vector
field than quadratic, the correlation functions of the type
$\langle\zeta_0 {\bf B}^2 {\bf B}^2\rangle$ only receives a
contribution from contractions  of the form
\beq \label{zBBBBint}
\langle\zeta_0(\vect{k}_1) {\bf B}^2(\vect{k}_2) {\bf B}^2(\vect{k}_3)\rangle = 4\int\frac{d^3 q_1}{(2\pi)^3} \int\frac{d^3 q_2}{(2\pi)^3}\langle\zeta(\vect{k}_1) B_i(\vect{k}_2-\vect{q}_1) B_j(\vect{k}_3-\vect{q}_2) \rangle\langle B_i(\vect{q}_1) B_j(\vect{q}_2)\rangle~,
\eeq

In order to evaluate this expression, we write $ \langle\zeta_0(\vect{k}_1) B_i(\vect{p}) B_j(\vect{r}) \rangle$ as the most general tensor function of $\vect{k}_1$, $\vect{p}$ and $\vect{r}$ with $\vect{p}=\vect{k}_2-\vect{q}_1$ and $\vect{r}=\vect{k}_3-\vect{q}_2$  (see appendix \ref{appzBB}) ,
\bea \label{zBBBBtensor}
\langle\zeta_0(\vect{k}_1) B_i(\vect{p}) B_j(\vect{r}) \rangle &=& (2\pi)^3\delta^{(3)}(\vect{k}_1+\vect{p}+\vect{r})
\left[A (\delta_{ij} \hat p\cdot \hat r -\hat r_{i}\hat p_{j})+D(\hat p\times\hat r)_i(\hat r\times\hat p)_j\right.\nonumber\\
& &\left. G((\hat p\times\hat r)_i(\hat p_{j}-\hat r_{j}\hat p\cdot \hat r) -(\hat p\times\hat r)_j(\hat r_{i}-\hat p_{i} \hat p\cdot \hat r))\right.\\
&&\left.+J(\hat p_{i}\hat p\cdot \hat r-\hat r_{i})(\hat p_{j}-\hat r_{j}\hat p\cdot \hat r)\right]P_{\zeta}(k_1)\sqrt{P_{\B}(p)P_{\B}(r)}
\eea
where $A$, $D$, $F$, and $J$ are general scalar functions of $\vect{k}_1$, $\vect{p}$, and $\vect{r}$.

The magnetic non-linearity parameter $b_{NL}$ is given by the trace of this tensor (see appendix \ref{appzBB}), and within this parametrization, the $D$, $F$, and the $J$ terms vanishes in the squeezed limit $k_1<< k_2,k_3$. This implies that in the squeezed limit, we can identify $A$ with $b_{NL}^{local}$ in the following way, $b_{NL}^{local}=-2A$.

In the most general case, one obtains
\bea \label{zBBBBgen}
\langle\zeta_0(\vect{k}_1) {\bf B}^2(\vect{k}_2) {\bf B}^2(\vect{k}_3)\rangle &=& 2(2\pi)^3\delta^{(3)}(\vect{k}_1+\vect{k}_2+\vect{k}_3)\int\frac{d^3 q}{(2\pi)^3} P_{\zeta}(k_1)P_{\B}(q)\sqrt{P_{\B}(p)P_{\B}(r)}\nonumber\\
&\times& \left[ A(\hat p\cdot \hat r+ (\hat p\cdot \hat q)( \hat r \cdot q)) +D((\hat p \times \hat r)\cdot(\hat r \times\hat p)) \right.\\
& & \left. +J((\hat p\cdot \hat r) -(\hat p\cdot \hat r)^3-((\hat q\cdot \hat p )(\hat p \cdot \hat r) -\hat q \cdot \hat r)(\hat q\cdot \hat p-(\hat q\cdot \hat r)( \hat p\cdot \hat r)))\right]~.\nonumber
\eea
Thus, from a computation of the correlation function of  the cross correlation of the vector mode with the curvature perturbation, $\langle\zeta_0(\vect{k}_1) A_i(\vect{k}_2) A_j(\vect{k}_3) \rangle $, in any specific model, we can then directly read of the coefficients $A,D,J$, as explained in the appendix \ref{appzBB}.

It is interesting to note that the symmetry arguments of \cite{Biagetti:2013qqa} can be used as a consistency check of the tensor structure of the leading logarithmic divergent contribution to these coefficients. The conformal symmetry of the future boundary of de Sitter space fixes the asymptotic tensor structure of (\ref{zBBBBtensor}), (\ref{zBBBBgen}), and in the case of scale invariant magnetic fields with $n=2$ one has for the leading logarithmically divergent term $A=-(\hat r\cdot \hat p)D$ and $G=J=0$, as is discussed in more details in appendix \ref{appzBB} and applied in the folded shape in (\ref{A&Bfolded}). However, in the squeezed limit, these leading logarithmic terms are suppressed by a factor of $k_1^3$, which vanishes in the exactly squeezed limit $k_1\to 0$, and in this limit  we can instead identify $A$ with $-b_{NL}^{local}/2$, which is obtained from the magnetic consistency relation \cite{Jain:2012ga,Jain:2012vm}, as mentioned above.

In section \ref{BMMzzbzb} we will carry out the angular integral and evaluate the correlation function in an explicit benchmark model. But for illustrative reasons, we consider some simplified shapes below. First we consider the case where (\ref {zBB1}) is maximal in the squeezed limit with a scale invariant $b_{NL}$, controlled by $A$ (as well as in the flat limit), and then we consider the case where (\ref {zBB1}) is maximal in the  orthogonal shape again with a scale invariant $b_{NL}$, which is controlled by D. Note that the $J$ term vanishes is these two limits, and describes shapes which interpolates between the squeezed or folded shape and the orthogonal shape.

\subsubsection*{Squeezed limit}

As mentioned above, in the squeezed limit the $D$, $F$, and the $J$
terms vanish, and due to momentum conservation we are lead to taking
also the squeezed limit of (\ref{zBBBBtensor}) under the integral.
Thus in the squeezed limit $k_1<< k_2, k_3$, we can evaluate
correlators of the form $\langle\zeta_0 B^2 B^2\rangle$ on
superhorizon scales as
  \beq
  \label{zbbbb_int}
  \langle\zeta_0(\k_1){\bf B}^2(\k_2){\bf B}^2(\k_3)\rangle_{\rm
  c}(t)
  =(2\pi)^3\delta(\k_1+\k_2+\k_3)AP_{\zeta}(k_1)\left(\frac{(aH)^{2n-4}}{\lambda}\right)^{2}C_B^2I(k_2,t)+5 {\rm
  p.} \ ,
  \eeq
where we can use that $b_{NL}^{local}=-2A$ in the squeezed limit.
Recall that $C_B$ denotes the amplitude of the magnetic spectrum
according to equation (\ref{P_B1}). The function $I(k,t)$ denotes a
momentum integral given by,
  \baq
  \label{I}
  I(k,t)&=&\int\frac{\d\q}{(2\pi)^3}
  q^{1-2n}|\k-\q|^{1-2n}\left(1+\frac{(\q\cdot(\k-\q))^2}{q^2|\k-\q|^2}\right)\Theta(aH-q)
  \Theta(q-k_0)\\\nonumber&&\times\Theta(aH-|\k-\q|)
  \Theta(|\k-\q|-k_0)\ .
  \eaq
As before, $aH$ is the horizon scale at the time $t$ when the
correlator is evaluated and $k_0\equiv a_0H_0$ is the horizon at the
time $t_0$ when we assume the generation of the magnetic fields
starts.

The integral can be computed analytically and for modes well outside
the horizon, $aH\gg k \gg a_0H_0$, the result is approximatively
given by
 \baq
 \label{Ieval}
 I(k,t)&=&\frac{k^{5-4n}}{4\pi^2}\left(C_1(n,k)+C_2(n,k)\left(\frac{k_0}{k}\right)^{4-2n}
 +C_3(n,k)\left(\frac{aH}{k}\right)^{5-4n}\right)\ .
 \eaq
Here we have defined the coefficients as
  \baq
  \label{C_i}
  C_{1}(n,k)&=&\frac{\sqrt{\pi}4^{2n-3}\Gamma(3-2n)}{(2n-3)\Gamma(7/2-2n)}\left(1+\frac{1}{{\rm cos}(2\pi n)}\right)
  \left(1+\frac{3+(4-2n)(2-2n)}{(1-2n)^2}\right)\\\nonumber
  C_{2}(n,k)&=&-\frac{16}{3(4-2n)}\left(1+\frac{18(2-n)}{5(3-n)}\left(\frac{k_0}{k}\right)^2
  +\frac{1026(2-n)}{35(4-n)}\left(\frac{k_0}{k}\right)^4
  +{\cal O}\left(\frac{2-n}{5-n}\left(\frac{k_0}{k}\right)^6\right)\right)\\\nonumber
  C_{3}(n,k)&=&\frac{4}{5-4n}\left(\rule{0pt}{4ex}\right.1-\frac{3(5-4n)}{8(3-4n)}\left(\frac{k}{aH}\right)^2
  +\frac{51(5-4n)}{640(1-4n)}\left(\frac{k}{aH}\right)^4
  +\frac{55(5-4n)}{21504(1+4n)}\left(\frac{k}{aH}\right)^6\\\nonumber
  &&
  +{\cal O}\left(\frac{5-4n}{3+4n}\left(\frac{k}{aH}\right)^8\right)\left)\rule{0pt}{4ex}\right.\ .
  \eaq
The result holds in the strong coupling regime $n>-1/2$. In the
coefficients $C_2$ and $C_3$ we have retained the higher order terms
which diverge for some values of $n$. The divergences cancel the
divergences of the constant $C_{1}$ for the corresponding values of
$n$ so that the full result (\ref{Ieval}) is finite.

In the scale-invariant case $n=2$, the result (\ref{Ieval}) reduces
to a simple logarithmic form. The squeezed limit correlator
$\langle\zeta B^2 B^2\rangle$ for the scale-invariant case is then
given by the expression
  \beq
  \label{zb2b2}
   \langle\zeta_0(\k_1){\bf B}^2(\k_2){\bf B}^2(\k_3)\rangle_{\rm c}
  =
  (2\pi)^3\delta(\k_1+\k_2+\k_3)\bnl^{local}P_{\zeta}(k_1)P_{\B}(k_2)\mathcal{P}_{\B}\left[\frac{5}{9}-\frac{4}{3}{\rm ln} \left(\frac{k_2}{k_0}\right)\right]+ 5 {\rm
  p.}
  \eeq
Note that here we have used the definitions $(2\pi^2){\cal
P}_{\zeta}(k)/k^3= P_{\zeta}(k)$ and $(2\pi^2){\cal P}_{B}(k)/k^3=
P_{B}(k)$.

\subsubsection*{Orthogonal shape}

Another simple example is case of a scale-independent $b_{NL}^{orthogonal}$. As evident from (\ref{bnlorth}), the $D$ term is related to the orthogonal shape where $\vect{k}_2\cdot\vect{k}_3=0$ in (\ref {zBB1}). In the case of $k_2 = k_3$, it follows from (\ref{bnlorth}) that,
\beq
D = b_{NL}^{orthogonal}
\eeq
while we will set $A=J=0$. With this ansatz, we can carry out the angular integrals in the scale-invariant limit  (for $n_B=0$), in order to obtain
 \baq
  \label{zb2b2flat}
   \langle\zeta_0(\k_1){\bf B}^2(\k_2){\bf B}^2(\k_3)\rangle_{\rm c}
  &=&
  (2\pi)^3\delta(\k_1+\k_2+\k_3)b_{NL}^{orthogonal}{P}_{\zeta}(k_1){P}_{\B}(k_2)\mathcal{P}_{\B}\\
  & &\times\frac{2}{3}\left[-1+{\rm ln}(2)+{\rm ln} \left(\frac{k_2}{k_0}\right)\right] + {\rm
  perm.}\nonumber
  \eaq

\subsection{Correlators of the form $\langle B^2 B^2 B^2\rangle$}

Correlators of this form have been examined in
\cite{Caprini:2009vk,Fujita:2013qxa} and we will make use of these
results. For $n>1/2$ the magnetic spectrum (\ref{P_B1}) is infrared
divergent. The convolution integral over a product of three
$P_B(q_i)$'s in $\langle B^2 B^2 B^2\rangle$ can then be
approximated by the contributions around the three poles of the
integrand. Assuming furthermore that all the wavenumbers are of
equal magnitude $k_i \sim k$, this gives the result
\cite{Fujita:2013qxa}
  \baq
  \label{b2b2b2}
  \langle{\bf B}^2(\k_1){\bf B}^2(\k_2){\bf B}^2(\k_3)\rangle_{\rm
  c}\Big|_{k_i\sim k}
  &=& (2\pi)^3\delta(\k_1+\k_2+\k_3)P_{B}(k)^2{\cal P}_B(aH)\times\\\nonumber&&
  \frac{1+{\rm cos}^2(\k_1,\k_2)}{6-3n}\left(\left(\frac{k}{aH}\right)^{4-2n}-\left(\frac{k_0}{aH}\right)^{4-2n}\right)+2{\rm p}.
  \eaq
Here $aH$ is the horizon scale at the time when the correlator is
evaluated and $k_0 = a_0H_0$ denotes the horizon scale as the
generation of magnetic fields started.

\section{Curvature perturbation induced by the magnetic fields}

The energy density of the magnetic fields is assumed to be small
during inflation. The magnetic fluctuations generated by the
inflationary expansion then amount to isocurvature perturbations
which seed the generation of adiabatic curvature perturbations. To
leading order in the coupling $\lambda$, the curvature perturbation
induced by the magnetic fields is given by
  \beq
  \label{zeta_iso}
  \zeta_{B}(t)=-\int_{t_0}^{t}dt\frac{H}{\rho+p}\left(\delta
  p_{B}-\frac{\dot{p}}{\dot{\rho}}\delta\rho_B\right)\ .
  \eeq
Here we have assumed that no curvature perturbation was generated by
the magnetic fields before the time $t_0$ so that
$\zeta_{B}(t_0)=0$. At any later time event $t>t_0$ the curvature
perturbation then consists of $\zeta_0$ generated independently of
magnetic fields and the induced contribution $\zeta_B$
  \beq
  \label{zetasplit}
  \zeta(t)=\zeta_{0}(t) +\zeta_{B}(t)\ .
  \eeq

Using that $\rho\simeq -p$ during inflation and that the magnetic
energy density is given by $\rho_{B}=\lambda B_i B_i/2 = 3 p_{B}$ we
can rewrite equation (\ref{zeta_iso}) in conformal time
$\eta=-1/(aH)$ as
  \beq
  \label{zetaB}
  \zeta_{B}(t)=\int_{\eta_0}^{\eta}d\ln \eta~\lambda(\eta)\frac{B_iB_i}{3H^2\epsilon}\
  .
  \eeq
Here $\lambda(\eta)$ is in general a time dependent quantity during
inflation corresponding to a non-minimal kinetic term for the vector
fields. For canonical vector fields $\lambda$ is one. In the
following we will neglect the time dependence of the Hubble rate $H$
and the slow roll parameter $\epsilon$ during inflation and treat
them as constants.

The induced curvature perturbation $\zeta_B$ will give two new contributions to the power spectrum of the primordial curvature perturbation. The two new contributions come from non-vanishing two point functions of the form $\left<\zeta_B \zeta_B\right>$ and  $\left<\zeta_0 \zeta_B\right>$.

Assuming a power law time dependence for the vector fields on
superhorizon scales (\ref{P_B1}) the spectrum of the induced
curvature perturbation from the  $\left<\zeta_B \zeta_B\right>$ correlation function is given by
  \baq
  \label{P_zB}
  {\cal P}_{\zeta_B}(\eta,k)&\simeq&\left(\frac{\Omega_B(\eta)}{\epsilon}\frac{4-2n}{1-(-k_0\eta)^{4-2n}}\right)^2\\\nonumber
  &&\times\left[
  \rule{0pt}{4ex}\right.\left(C_1(n,k)+C_2(n,k)\left(\frac{k_0}{k}\right)^{4-2n}\right)
  \left(\frac{1-(-k\eta)^{4-2n}}{4-2n}\right)^2\\\nonumber
  &&+2C_3(n,k)\left(\frac{1}{3(4-2n)}-\frac{1}{(4-2n)(2n-1)}\left(-k\eta\right)^{4-2n}
  +\frac{1}{3(2n-1)}\left(-k\eta\right)^{3}\right)
  \left] \rule{0pt}{4ex}\right.\ ,
  \eaq
where the coefficients $C_i$ are given by equation (\ref{C_i}). The
energy density of the magnetic fields, $\rho_B=3 H^2 \Omega_{B}$ is
given by
  \baq
  \langle\rho_B(\eta)\rangle&=&\frac{\lambda}{2a^4}\langle((\partial_iA_j(t,\x))(\partial_iA_j(t,\x))\rangle\\
  &=& \frac{C_B}{4\pi^2(4-2n)}\left(1-\left(\frac{k_0}{aH}\right)^{4-2n}\right)\ .
  \eaq

The contribution to the induced power spectrum from the  $\left<\zeta_0 \zeta_B\right>$ correlation function, can be evaluated using (\ref{zBB1}) and assuming that $b_{NL}$ is momentum independent, which is equivalent to the local ansatz. One then finds
 \bea \label{P_zBbnl}
   {\cal P}_{\zeta_B}^{(b_{NL})}(\eta,k)&\simeq&\frac{1}{2}b_{NL}  {\cal P}_{\zeta}(k)\left(\frac{\Omega_B(\eta)}{\epsilon}\frac{4-2n}{1-(-k_0\eta)^{4-2n}}\right)\times\nonumber\\
 && \frac{1}{4-2n}\left(\log\left(\frac{k_0}{aH}\right)-   \frac{1}{4-2n}\left(\frac{k_0}{aH}\right)^{4-2 n}\right)~.
   \eea

\subsection{The induced bispectrum amplitudes}

The curvature perturbation induced by the magnetic fields is
quadratic in the fluctuations of the vector field
$A_i$ and hence obeys a non-Gaussian statistics. The generation of
magnetic fields may therefore significantly affect the three point
function of primordial correlators as well as higher order
non-Gaussian statistics. Schematically, the three point correlator
of primordial perturbations takes the form
\beq
\langle\zeta\zeta\zeta\rangle =\langle\zeta_0\zeta_0\zeta_0\rangle
+3\langle\zeta_0\zeta_0\zeta_B\rangle +3
\langle\zeta_0\zeta_B\zeta_B\rangle +
\langle\zeta_B\zeta_B\zeta_B\rangle\ .
\eeq
In the canonical slow roll inflation, the first term represents the
pure inflaton contribution and is slow roll suppressed. The
magnetically induced terms may however be sizeable, depending on the
model. Using equation (\ref{zetaB}) these can be written
respectively as
  \baq
  \label{z0z0zb_int}
  \langle\zeta_0(\k_1)\zeta_0(\k_2)\zeta_B(\k_3)\rangle(\eta)&=&\frac{1}{3H^2\epsilon}
  \hspace{-7pt}\int\limits_{-1/k_3}^{\eta}\hspace{-7pt}d\eta'\frac{\lambda(\eta')}{\eta'}
  \langle\,\zeta_0(\k_1,\eta)\zeta_0(\k_2,\eta){\bf B}^2(\k_3,\eta')\,\rangle_{\rm c}\
  ,\\
  \label{z0zbzb_int}
  \langle\zeta_0(\k_1)\zeta_B(\k_2)\zeta_B(\k_3)\rangle(\eta)&=&\left(\frac{1}{3H^2\epsilon}\right)^2\hspace{-7pt}
  \int\limits_{-1/k_2}^{\eta}\hspace{-7pt}\d\eta'
  \hspace{-7pt}\int\limits_{-1/k_3}^{\eta}\hspace{-7pt} d\eta''\frac{\lambda(\eta')\lambda(\eta'')}{\eta'\eta''}\times\\\nonumber
 & &  \langle\,\zeta_0(\k_1,\eta){\bf B}^2(\k_2,\eta'){\bf B}^2(\k_3,\eta'')\,\rangle_{\rm c}\
  ,\\
  \label{zbzbzb_int}
  \langle\zeta_B(\k_1)\zeta_B(\k_2)\zeta_B(\k_3)\rangle(\eta)&=&\left(\frac{1}{3H^2\epsilon}\right)^3\hspace{-5pt}
  \int\limits_{-1/k_1}^{\eta}\hspace{-7pt}\d\eta'
  \hspace{-7pt}\int\limits_{-1/k_2}^{\eta}\hspace{-7pt}d\eta''
  \hspace{-7pt}\int\limits_{-1/k_3}^{\eta}\hspace{-7pt}d\eta'''\frac{\lambda(\eta')\lambda(\eta'')\lambda(\eta''')}
  {\eta'\eta''\eta'''}\times\\\nonumber&&
  \langle\,{\bf B}^2(\k_1,\eta'){\bf B}^2(\k_2,\eta''){\bf B}^2(\k_3,\eta''')\,\rangle_{\rm c}\
  .
  \eaq
During inflation fluctuations of magnetic fields amount as
isocurvature perturbations and the total curvature perturbation
$\zeta$ therefore keeps evolving on superhorizon scales. As the
magnetic energy density scales as radiation the isocurvature
perturbations induced by magnetic fields vanish as the universe
becomes radiation dominated after the end of inflation and $\zeta$
freezes to a constant value. We are therefore interested in
evaluating the curvature perturbation $\zeta$ and its correlators at
the beginning of the radiation era which we assume coincides with
the end of inflation. In the following we will thus set
$\eta=\eta_{\rm end}$.

Using the power law assumption for the magnetic spectrum
(\ref{P_B1}), the correlator (\ref{z0z0zb_int}) can be written as
 \baq
 \label{zzzb_step1}
 \langle\zeta_0(\k_1)\zeta_0(\k_2)\zeta_B(\k_3)\rangle(\eta)&=&\frac{1}{3H^2\epsilon}
  \hspace{-7pt}\int\limits_{-1/k_3}^{\eta_{\rm end}}\hspace{-7pt}d\eta'\frac{\lambda(\eta)}{\eta'}
  \left(\frac{\eta'}{\eta}\right)^{4-2n}\times
  \\\nonumber
  &&
  \langle\,\zeta_0(\k_1)\zeta_0(\k_2)\int{\bf B}(\q){\bf B}(\k_3-\q)\,\rangle_{\rm
  c}(\eta)\times\\\nonumber
  &&\theta(q-k_0)\theta(|\k_3-\q|-k_0)\theta(-1/\eta'-q)\theta(-1/\eta'-|\k_3-\q|)\ .
\eaq
Assuming the local Ansatz (\ref{Blocal}) for magnetic fields, the
equal time correlator on the right hand side of (\ref{zzzb_step1})
is given by equation (\ref{zzbb_sol}) after changing the limits
integral in (\ref{zzbb_sol}) to match with those above. Inserting
the expression (\ref{zzbb_sol}) and performing the time integral we
then arrive at the result
  \baq
  \label{Bz0z0zb}
  \langle\zeta_0(\k_1)\zeta_0(\k_2)\zeta_B(\k_3)\rangle
  &\simeq&
  (2\pi)^3\delta(\k_1+\k_2+\k_3)P_{\zeta}(k_1)P_{\zeta}(k_2)\frac{\Omega_{B}}{\epsilon}\times\\\nonumber
&&  \left(\frac{1}{2}(\bnl^{\rm local})^2+\frac{2}{3}\cnl^{\rm
  local}\right)\times\\\nonumber
  &&
  \frac{2}{1-e^{(2n-4)N_0}}
  \left(N_3+
  \frac{e^{(2n-4)N_0}}{2n-4}\left(1-e^{-(2n-4)N_3}\right)\right)
  +2{\rm p.}\ .
  \eaq
Here $N_0={\rm ln}(a_{\rm end}/a_0)$ denotes the number of
e-foldings from the onset of magnetic field generation $a_0$ to the
end of inflation and $N_i={\rm ln}(a_{\rm end}/a_{k_i})$ the number
of e-foldings from the horizon exit of the mode $k_i$.

It is conventional to parameterize the three point function by the
parameter $\fnl$ measuring the bispectrum amplitude normalized by
the square of the spectrum, which is defined in terms of the
bispectrum
\beq
\left< \zeta(\vect{k}_1)\zeta(\vect{k}_2)\zeta(\vect{k}_2)\right> \equiv (2\pi)^3\delta^{(3)}(+\vect{k}_1+\vect{k}_2+\vect{k}_3)B_{\zeta}(\vect{k}_2,\vect{k}_3,\vect{k}_1)
\eeq
by
\beq
B_{\zeta}\equiv \frac{6}{5}f_{NL}\sum_{a<b}P_{\zeta}(k_a)P_{\zeta}(k_b)~.
\eeq

The induced $\fnl$ generated by the
correlator $\langle\zeta\zeta\zeta_B\rangle$ is then given by
  \baq
  \label{fzzb1}
  \fnl^{\zeta\zeta\zeta_{B}}&=&-\frac{5}{18}\left(3(\bnl^{\rm local})^2+4\cnl^{\rm local}\right)\frac{\Omega_B}{\epsilon}
  \frac{1}{1-e^{(2n-4)N_0}}\times
  \\\nonumber
  &&
  \left(k_3^3N_3+
  \frac{k_3^3e^{(2n-4)N_0}}{2n-4}\left(1-e^{-(2n-4)N_3}\right)\right)
  \times\left(k_1^3+k_2^3+k_3^3\right)^{-1}\ .
  \eaq

In a similar way, assuming the local Ansatz (\ref{Blocal}) and using
equations (\ref{P_B1}), (\ref{zbbbb_int}) and (\ref{Ieval}) in
(\ref{z0zbzb_int}) we find the non-linearity parameter associated to
the correlator $\langle\zeta_0\zeta_B\zeta_B\rangle$ given by
  \baq
  \label{fzbb1}
  \fnl^{\zeta\zeta_{B}\zeta_{B}}&=&
  \frac{5}{6}\frac{\bnl^{\rm
  local}}{{\cal P}_{\zeta}}\left(\frac{\Omega_B}{\epsilon}\frac{4-2n}{1-e^{(2n-4)N_0}}\right)^2
  \times\\\nonumber
  &&
  \left[
  \rule{0pt}{4ex}\right.
  k_3^3\left(C_1(n,k)+C_2(n,k)e^{(2n-4)(N_0-N_2)}\right)e^{(2n-4)(N_2-N_3))}
  \left(\frac{(1-e^{(2n-1)N_2})(1-e^{(2n-1)N_3})}{(4-2n)^2}\right)
  \\\nonumber
  &&
  +C_3(n,k)k_3^3\left(1+e^{(2n-1)(N_3-N_2)}
  \rule{0pt}{3ex}\right.\left)
  \rule{0pt}{3ex}\right.\frac{1}{(2n-1)(4-2n)}\left(
  \rule{0pt}{3ex}\right.
  1-e^{(2n-4)N_2}\left)
  \rule{0pt}{3ex}\right.\\\nonumber
  &&+C_3(n)
  \frac{2k_3^3}{3(2n-1)}(1+e^{-3 N_{2}})+5 {\rm p.}
  \left]\rule{0pt}{4ex}\right.
  \times
  \left[
  k_1^3+k_2^3+k_3^3
  \right]^{-1}
  \ .
  \eaq
Here the coefficients $C_i$ are given by equation (\ref{C_i}).

Finally, using the expression for magnetic spectrum (\ref{P_B1}) in
(\ref{zbzbzb_int}) and performing the integrals one obtains for
nearly equilateral configurations $k_i\sim k$ the result
\cite{Fujita:2013qxa} (see also \cite{Caprini:2009vk})
  \baq
  \fnl^{\zeta_B\zeta_B\zeta_B}&\simeq&-\frac{1}{{\cal P}_{\zeta}^2}\left(
 \frac{\Omega_B}{\epsilon}\right)^3 \frac{5}{9(4-2n)}\left(1-e^{(2n-4)(N_k-N_0)}\right)
  \left(\frac{1-e^{(2n-4)N_k}}{1-e^{(2n-4)N_0}}\right)^3\times\\\nonumber
  &&\frac{8}{3}\left(1+{\rm cos}^2(\k_1,\k_2)+2{\rm
  p.}\right)\ .
  \eaq
This expression gives the non-linearity parameter
$\fnl^{\zeta_B\zeta_B\zeta_B}$ measuring the amplitude of the
induced bispectrum of the form\footnote{The trispectrum induced by Gaussian magnetic fields were computed in \cite{Trivedi:2011vt}.}
$\langle\zeta_B\zeta_B\zeta_B\rangle$.

\subsection{Observational constraints}

The results of the Planck satellite place stringent constraints on
the primordial non-Gaussianity. These bounds can be used to place
constraints on the magnetic non-linearity parameters
$b_{NL}^{local}$ and $c_{NL}^{local}$ in terms of equations
(\ref{fzzb1}) and (\ref{fzbb1}). In this way the Planck constraints
can open interesting new insights on the gauge field couplings in
the early universe.

Here we will exemplify the resulting constraints on $b_{NL}^{local}$
and $c_{NL}^{local}$ concentrating on the case of flat magnetic
fields $n=2$ only. In this limit the spectrum of curvature
perturbations (\ref{P_zB}) induced by the magnetic fields is given
by
  \baq
  {\cal P}_{\zeta_B}=\frac{16}{3}\left(\frac{\Omega_{B}}{\epsilon}\right)^2\left(\frac{N_{\rm
  CMB}}{N_0}\right)^2(N_0-N_{\rm CMB})\ ,
  \eaq
while from (\ref{P_zBbnl}) we obtain the additional contribution
\beq\label{P_zbbnl2}
  {\cal P}_{\zeta_B}^{(b_{NL})}=\frac{1}{2}b_{NL}\left(\frac{\Omega_{B}}{\epsilon}\right) {\cal P}_{\zeta}~ N_0\ .
\eeq

The energy density of magnetic fields at the time of inflation
$\Omega_B$ can be related to the amplitude of magnetic fields today
using
  \beq\label{Omega}
  \Omega_{B} \sim 10^{-7}\left(\frac{B_{\rm today}}{10^{-9} {\rm
  G}}\right)^2\ ,
  \eeq
where we have used that the magnetic energy density scales as
radiation and that the radiation energy density today is given by
$\rho_{\rm rad.}\sim 10^{-51}$ GeV${}^4$. Using this we can then
express the spectrum in the form
  \baq
  \label{P_zb_n2_btoday}
  {\cal P}_{\zeta_B}\sim 10^{-10}\left(\frac{B_{\rm today}}{10^{-9}{\rm
  G}}\right)^4\left(\frac{0.01}{\epsilon}\right)^2\left(\frac{N_{\rm
  CMB}}{N_0}\right)^2(N_0-N_{\rm CMB})\ .
  \eaq

The direct magnetic field constraints by Planck set the bound
$B_{\rm today}\lesssim 10^{-9}$ G \cite{Ade:2013zuv} on Mpc scales.
As can be seen in equation (\ref{P_zb_n2_btoday}), the indirect
constraint from amplitude of induced curvature perturbation ${\cal
P}_{\zeta_B}\leqslant {\cal P}_{\zeta}=2.44\times 10^{-9}$
\cite{Ade:2013zuv} is comparable \cite{Bonvin:2013tba} and can even
be tighter if $\epsilon\ll 0.01$ or the generation of magnetic
fields started long before the horizon exit of observable modes. For
a further discussion of this latter point see the end of this
section.

The contribution $P_{\zeta_B}^{(b_{NL})}$ has not been considered before. From ({\ref{P_zbbnl2}) and (\ref{Omega}), we find
\beq\label{P_zbbnl3}
  {\cal P}_{\zeta_B}^{(b_{NL})}\sim\frac{1}{2}b_{NL} 10^{-5}\left(\frac{B_{\rm today}}{10^{-9}{\rm
  G}}\right)^2 \left(\frac{0.01}{\epsilon}\right) {\cal P}_{\zeta}~ N_0\ .
\eeq
Assuming $B_{\rm today}\sim 10^{-9}$ G and $N_0\sim 60$ leads to an
upper bound $b_{NL}\lesssim 10^{4}$. If $b_{NL}$ is larger, it will
imply a stronger upper bound on the magnetic field today,  $B_{\rm
today}$, than the model independent bound inferred from
(\ref{P_zb_n2_btoday}).

For the flat spectrum and in the the squeezed limit $k_1\ll k_2\sim
k_3$ the induced non-linearity parameter of type
$\fnl^{\zeta\zeta\zeta_B}$ (\ref{fzzb1}) becomes
\beq
\fnl^{\zeta\zeta\zeta_B}\simeq \frac{5}{9}\left(3(\bnl^{\rm
local})^2+4\cnl^{\rm local}\right)\frac{\Omega_B}{\epsilon} N_{\rm
CMB}\ ,
\eeq
where $N_{\rm CMB}$ denotes the number of e-foldings from the
horizon exit of observable modes. Expressing the non-linearity
parameter in terms of the magnetic field amplitude today and using
that $N_{\rm CMB}\sim 60$, we find
\beq
\label{fzzb_n2_btoday}
\fnl^{\zeta\zeta\zeta_B}\sim 3\times 10^{-4} \left(3(\bnl^{\rm
local})^2+4\cnl^{\rm local}\right)\left(\frac{B_{\rm today}}{10^{-9}
{\rm  G}}\right)^2\left(\frac{0.01}{\epsilon}\right)\ .
\eeq

This should be contrasted with the Planck constraint on local
non-Gaussianity $-8.9<\fnl<14.3$ (95\% C.L.) \cite{Ade:2013ydc}. The
resulting bounds on the magnetic non-linearity parameters $\bnl$ and
$\cnl$ are illustrated in Figure \ref{fzzb_bounds}.
\begin{figure}[!h]
\begin{center}
\includegraphics[width=7.5cm]{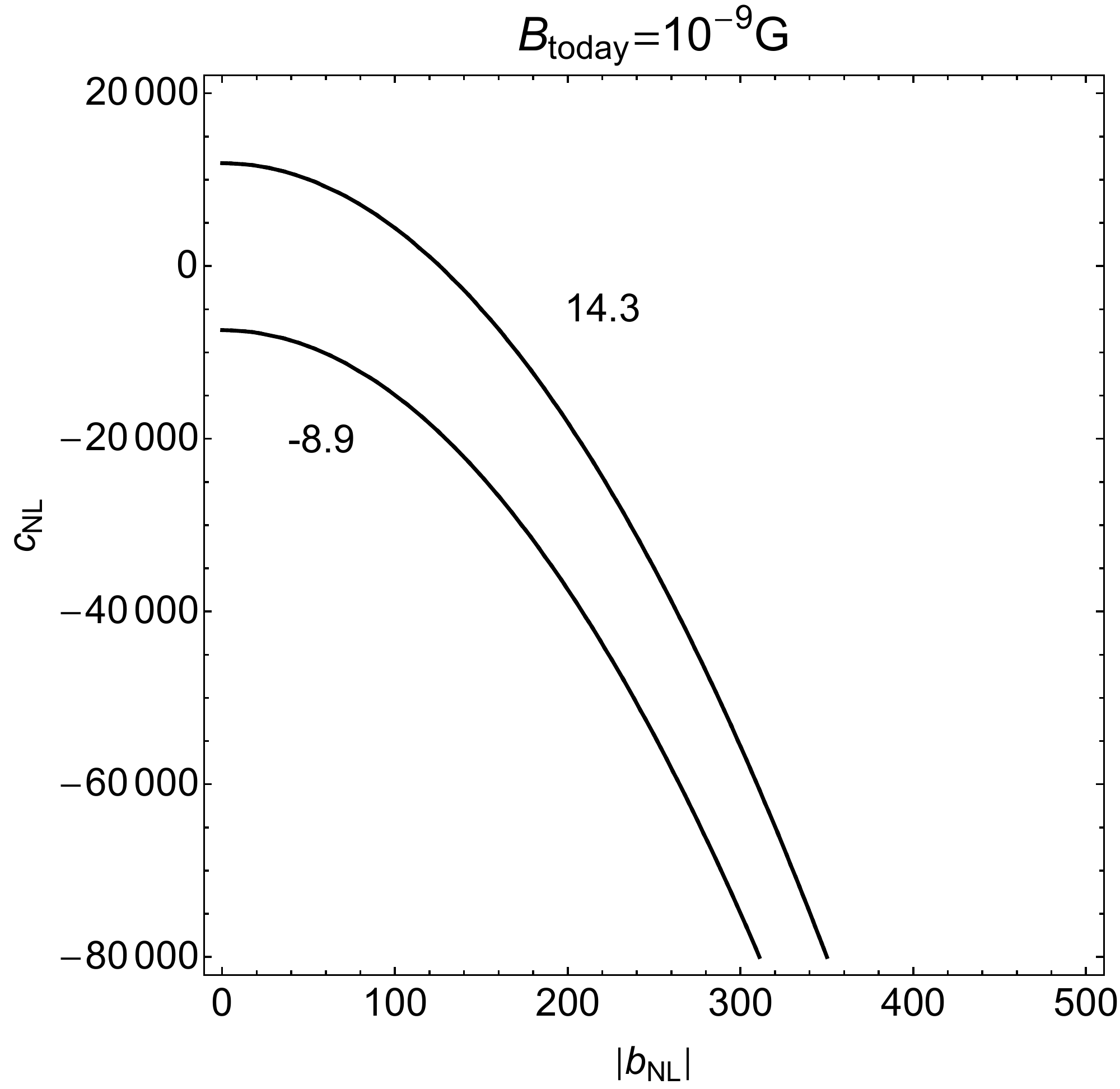}
\includegraphics[width=7.5cm]{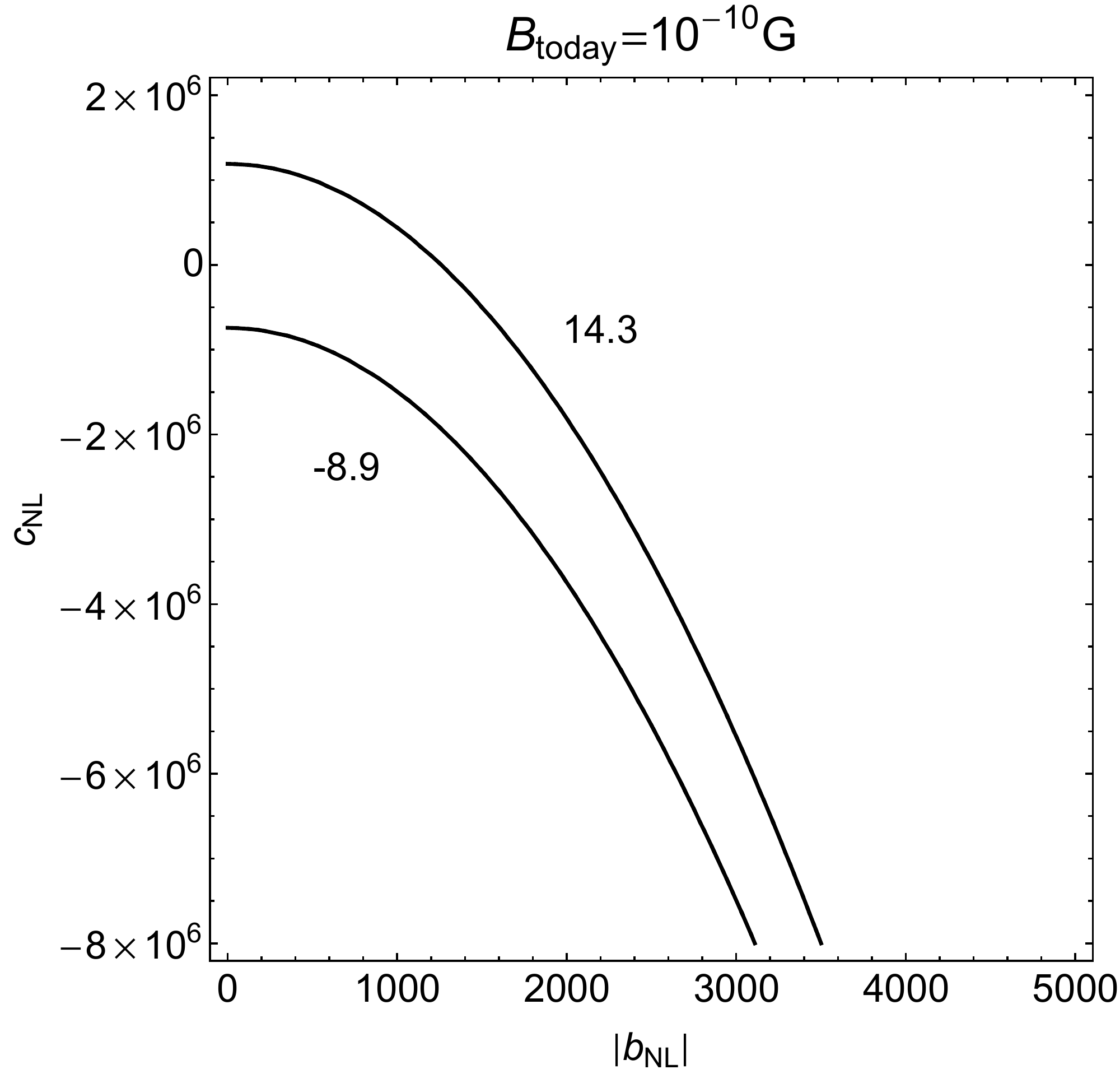}

\end{center}
\caption{The induced non-Gaussianity parameter
$\fnl^{\zeta\zeta\zeta_B}$ as a function of $\bnl$ and $\cnl$. The
regime compatible with the non-detection by Planck lies between the
contours $\fnl=-8.9$ and $\fnl=14.3$, bounding the Planck $2\sigma$
for $\fnl$. The left panel corresponds to situation $B_{\rm today} =
10^{-9}$ G marginally allowed by observations and corresponding to
${\cal P}_{\zeta_B}\sim {\cal P}_{\zeta}$ and the right panel
depicts the case for $B_{\rm today} = 10^{-10}$ G where the magnetic
contribution to curvature perturbations is subdominant. Here we have
set the inflationary slow roll parameter $\epsilon=0.01$.}
\label{fzzb_bounds}
\end{figure}
If the magnetic field amplitude is close to the observational upper
bound $B_{\rm today}\sim 10^{-9}$ G one obtains a tight constraint
$|\bnl|\lesssim 10^2$, barring cancellations against the parameter
$\cnl$. In this case the magnetic contribution to the amplitude of
curvature perturbations (\ref{P_zb_n2_btoday}) is also
non-negligible. For smaller magnetic field amplitudes the bounds on
$\bnl$ and $\cnl$ get relaxed as the the induced non-Gaussianity
(\ref{fzzb_n2_btoday}) scales as $\fnl^{\zeta\zeta\zeta_B}\propto
B^2$.

In a similar way, in the flat case and squeezed limit the induced
non-linearity parameter of type $\fnl^{\zeta\zeta_B\zeta_B}$
(\ref{fzbb1}) takes the form
  \beq
  \fnl^{\zeta\zeta_B\zeta_B} \simeq 0.2 \bnl \left(\frac{B_{\rm today}}{10^{-9}
{\rm  G}}\right)^2\left(\frac{0.01}{\epsilon}\right)^2
\left(\frac{N_{\rm
  CMB}}{N_0}\right)^2(N_0-N_{\rm CMB})\ .
   \label{fzzbzb_n2_btoday}
  \eeq
Here we have used that ${\cal P}_{\zeta}=2.44\times 10^{-9}$
\cite{Ade:2013zuv} and $\Omega_{B} \sim 10^{-7} B^2_{\rm today} {\rm
nG}^{-2}$. Comparing with the result (\ref{fzzb_n2_btoday}) we find
that the induced non-Gaussianity of type
$\fnl^{\zeta\zeta_B\zeta_B}$ generically yields a stronger
constraint on $\bnl$ but the result depends on the duration of the
epoch when magnetic fields were generated. Setting conservatively
$N_0-N_{\rm CMB}=5$ and choosing $N_{\rm CMB}=60$ one obtains the
constraints depicted in Figure \ref{fzbb_bounds}.
\begin{figure}[!h]
\begin{center}
\includegraphics[width=8cm]{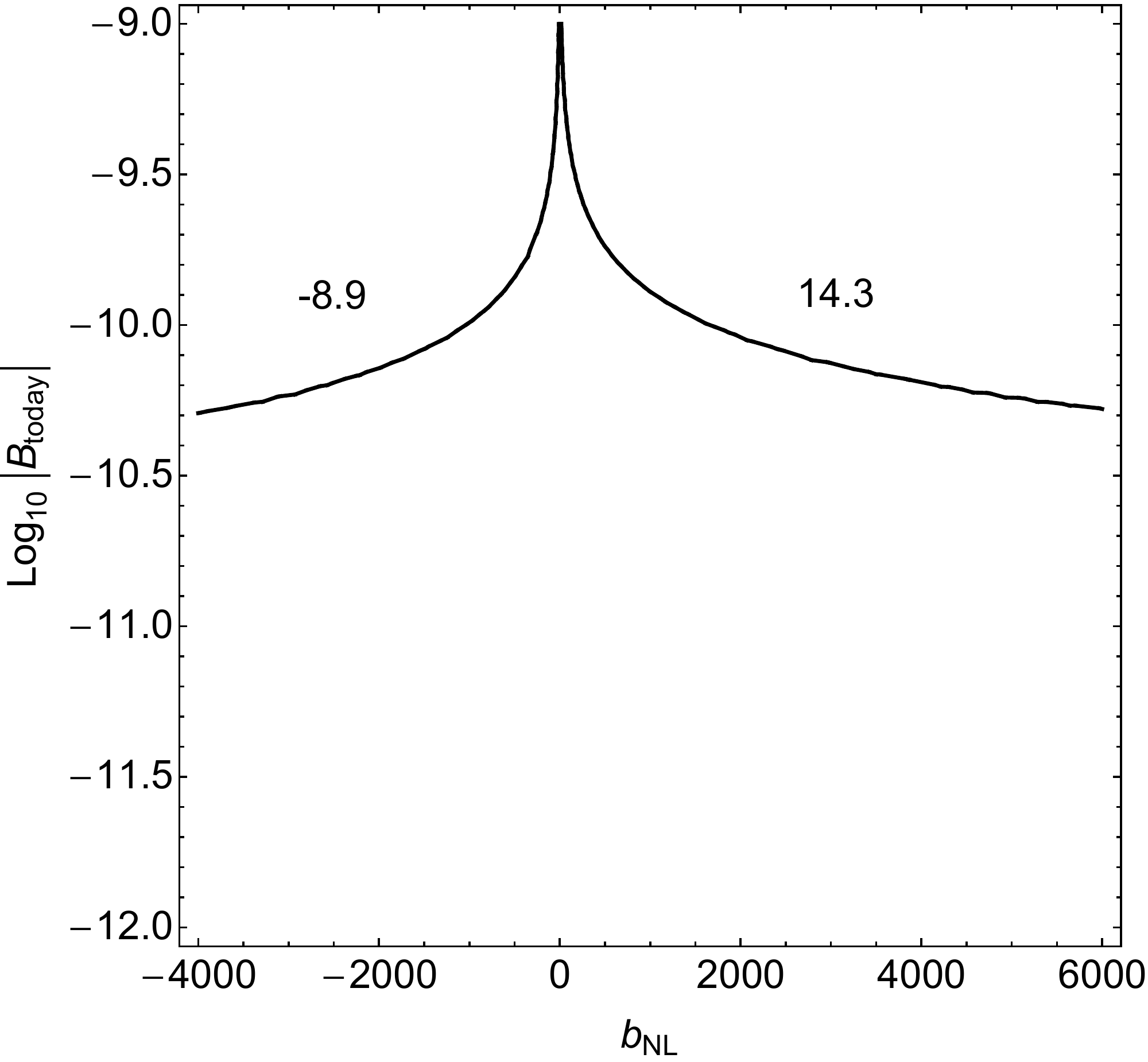}
\end{center}
\caption{The induced non-Gaussianity parameter
$\fnl^{\zeta\zeta_B\zeta_B}$ as a function of $\bnl$ and $B_{\rm
today}$. The regime compatible with the non-detection by Planck lies
below the contours $\fnl=-8.9$ and $\fnl=14.3$, corresponding to the
$2\sigma$ region for $\fnl$. Here we have set the inflationary slow
roll parameter $\epsilon=0.01$ and assumed the magnetic field
generation started $5$ e-foldings before the horizon exit of largest
observable modes.}
 \label{fzbb_bounds}
\end{figure}

We reiterate that our result assumes the generation of magnetic
fields started at some time $t_0$ corresponding to the horizon scale
$k_0=a_0H_0$ and we assume this was before the horizon crossing of
largest observable modes $t_0<t_{\rm CMB}$. Formally we are then
studying the statistics of fluctuations in a patch of size
$k_0^{-1}$ which does not in general correspond to the statistics
which can be measured in the observable patch of size $k_{\rm
CMB}^{-1}<k_0^{-1}$, see
\cite{Giddings:2010nc,Byrnes:2010yc,Giddings:2011zd,Byrnes:2011ri,Nurmi:2013xv,Byrnes:2013qjy,Nelson:2012sb,LoVerde:2013xka,LoVerde:2013dgp}.
If $k_{0}\ll k_{\rm CMB}$ the long-wavelength fluctuations of
magnetic fields generate an effective background field for our patch
and we should instead consider the statistics of fluctuations around
this background. In order to avoid these complications here, we
restrict to the case where $N_0-N_{\rm CMB}\lesssim {\cal O}(10)$ so
that difference of the statistics of fluctuations in the patches
$k_{0}^{-1}$ and $k_{\rm CMB}^{-1}$ is small unless the curvature
perturbation would be highly non-Gaussian \cite{Nurmi:2013xv}. This
approach remains valid even if the generation of magnetic fields
would have started long before the horizon exit of our observable
modes but then implicitly assumes that patch $k_0^{-1}$ occupies a
region where the effective background magnetic field vanishes.

\section{Benchmark model}

In order to estimate the natural values for $b_{NL}$ and $c_{NL}$,
we consider the non-Gaussianities generated by amplification of
magnetic fields during inflation in a specific model. We assume the Lagrangian is of the
form
  \beq
  {\cal
  L}=\frac{1}{2}R-\frac{1}{2}\partial^{\mu}\phi\partial_{\mu}\phi-V(\phi)-\frac{1}{4}\lambda(\phi)F_{\mu\nu}F^{\mu\nu}\
  ,
  \eeq
where $\lambda(\phi)$ takes a power law form
\cite{Turner:1987bw,Ratra:1991bn,Dolgov:1993vg,Davis:2000zp,Giovannini:2002ki,Bertolami:2005np,Bamba:2006ga}
  \beq
  \lambda=\lambda_{I}\left(\frac{a}{a_I}\right)^{2n} \ ,
  \eeq
and $a_I$ denotes the end of inflation. We assume $\phi$ is a slowly
rolling scalar field and consider fluctuations around the
homogeneous FRW background solution,
  \beq
  \bar{\phi}=\bar{\phi}(t)\ ,\qquad {\bar A}_{\mu}=0\ .
  \eeq
We concentrate on the exponent values $n>-1/2$ for which the large
scale modes of the vector potential $A_i\sim a^n/\lambda^{1/2}$ are
nearly constant and the backreaction of the magnetic fields to the
inflationary dynamics can be kept small. But as already mentioned,
the results in the other regime can be obtained by use of the
electromagnetic duality, which leaves the result invariant under a
simultaneous exchange of the electric and magnetic field and $n\to
-n$  \cite{Buonanno:1997zk,Brustein:1998kq,Giovannini:2009xa}.

The electromagnetic part of the action for perturbations is
quadratic in $A_{\mu}$, the fluctuation of the vector potential
around the zero background, including terms schematically of the
form
  \beq
  {\cal L}_{\rm pert.}\supset A^2\zeta^m\ .
  \eeq
Here $\zeta$ is the curvature perturbation and $m$ is a positive
integer. In the Coulomb gauge the magnetic field is related to the
vector potential as $B_{i}=a^{-2}\epsilon_{ijk}\partial_jA_k$.

On superhorizon scales $k\ll aH$, and treating the Hubble rate
during inflation as a constant $H=H_I$, the spectrum of the magnetic
fields generated during inflation then takes the form
  \beq
  \label{PBas}
  P_{\B}(\eta,k)=\frac{4^n\Gamma^2(n+1/2)}{\lambda(\eta)\pi}\frac{H_I^4}{k^3}(-k\eta)^{4-2n}\ .
  \eeq
Due to the fact that the vector potential is approximately constant, the
energy density of the electromagnetic field is dominated by the
magnetic part
  \baq
  \langle\rho_B(\eta)\rangle&=&\frac{\lambda}{2a^4}\langle((\partial_iA_j(t,\x))(\partial_iA_j(t,\x))\rangle\\
  &=&\frac{1}{2}\int\frac{dq^3}{(2\pi)^3}P_B(\eta,q)\theta(q-k_0)\theta(-1/\eta-q)\\
  &=& \frac{4^{n-1}\Gamma^2(n+1/2)H_I^4}{\pi^3(4-2n)}\left(1-e^{(2n-4)N_0}\right)\ .
  \eaq
Here $k_0$ corresponds to the largest scale at which inflationary
magnetic fields are generated. The contribution of the magnetic
fields to the total energy density during inflation is then
controlled by
  \beq
  \frac{\langle\rho_B(\eta)\rangle}{\rho_{\rm tot.}}=r_T{\cal
  P}_{\zeta}\frac{4^{n-1}\Gamma^2(n+1/2)}{6\pi(4-2n)}\left(1-e^{(2n-4)N_0}\right)\
  ,
  \eeq
where $r_T=16\epsilon\lesssim 0.02$ is the tensor to scalar ratio.
Using that ${\cal P}_{\zeta}=2.4\times 10^{-9}$ and requiring that
the magnetic fields remain subdominant for at least a period of 60
e-foldings one obtains the constraint $n\lesssim 2.2$ \cite{Demozzi:2009fu,Ferreira:2013sqa}, unless the scale of inflation is very low \cite{Ferreira:2013sqa}.

The energy density of the magnetic fields sources the generation of
adiabatic curvature perturbation according to the formula
(\ref{zeta_iso}). As the magnetic spectrum (\ref{PBas}) is of the
form (\ref{P_B1}) the spectrum of induced curvature perturbation
$\zeta_B$ is directly obtained from equation (\ref{P_zB}) by
substituting the corresponding value of $C_B$. This yields the
result
  \baq
  {\cal P}_{\zeta_B}(\eta,k)&\simeq&{\cal
  P}_{\zeta}^2\left(\frac{4^{n+1}\Gamma^2(n+1/2)}{6\pi}\right)^2\times\\\nonumber
  &&\left[
  \rule{0pt}{4ex}\right.\left(C_1(n,k)+C_2(n,k)e^{(2n-4)(N_0-N_{\rm CMB}}\right)
  \left(\frac{1-e^{(2n-4)N_{\rm CMB}}}{4-2n}\right)^2\\\nonumber
  &&+2C_3(n,k)\left(\frac{1}{3(4-2n)}-\frac{1}{(4-2n)(2n-1)}e^{(2n-4)N_{\rm CMB}}
  +\frac{1}{3(2n-1)}e^{3N_{\rm CMB}}\right)
  \left] \rule{0pt}{4ex}\right.\ ,
  \eaq
where the coefficients $C_i$ are given by equation (\ref{C_i}).

In the limit of a flat spectrum for the magnetic fields, $n=2$, the
leading part of the result takes the simple logarithmic form
  \beq
  \label{PzetaBflat}
  {\cal P}_{\zeta_B}^{n=2}(\eta,k)=
  192\,{\cal P}_{\zeta}^2N_{\rm CMB}^2(N_0-N_{\rm CMB})\ .
  \eeq
in agreement with \cite{Barnaby:2012tk}. For a discussion of this apparently coincidental agreement, see appendix \ref{source}.

Similarly, the $b_{NL}$ dependent contribution to the power spectrum gives in the flat limit
\beq\label{P_zbbnl4}
 {\cal P}_{\zeta_B}^{(b_{NL}) n=2}(\eta,k)=
  3 \,b_{NL}\,{\cal P}_{\zeta}^2 N_0^2\ .
  \eeq
We notice that for moderate values of $b_{NL}$, the new contribution
to the power spectrum (\ref{P_zbbnl4}) is the dominant one, although
in the local approximation where $b_{NL} = -4$, the contribution
(\ref{PzetaBflat}) is larger.

\subsection{Induced bispectrum $\langle\zeta_0\zeta_0\zeta_B\rangle$}\label{BMMzzzb}

In the squeezed limit the amplitude of the correlator
$\langle\zeta_0\zeta_0\zeta_B\rangle$ can be directly obtained from
equation (\ref{fzzb1}) using the value of $C_B$ obtained by
comparing the expressions (\ref{P_B1}) and (\ref{PBas}).
This yields the result
  \baq
  \label{fzzb}
  \fnl^{\zeta\zeta\zeta_{B}}&=&-20 n^2{\cal
  P}_{\zeta}\frac{4^{n+1}\Gamma^2(n+1/2)}{18\pi}\times\\\nonumber
  &&\frac{k^3_3}{4-2n}\left(N_3+
   \frac{e^{(2n-4)N_0}}{4-2n}\left(1-e^{-(2n-4)N_3}\right)+2{\rm p.}
  \right)\times\left(k_1^3+k_2^3+k_3^3\right)^{-1}\ .
  \eaq

The non-detection of primordial bispectrum by Planck translates into
constraints on the parameters $\bnl$ and $\cnl$. Their values for
the benchmark model have been computed in appendix \ref{appPB}. In
the limit of a flat spectrum of the magnetic fields $n=2$, the
induced bispectrum (\ref{Bz0z0zb}) takes a local shape and the
momentum dependence of the induced $\fnl$ vanishes
  \beq
  \label{fzzb_flat}
  \fnl^{\zeta\zeta\zeta_{B},n=2}=-80{\cal
  P}_{\zeta}N_0^2\ .
  \eeq
Thus, in the local approximation the induced bispectrum from $\left<\zeta \zeta B^2\right>$ is well within the observational limits  $-8.9<\fnl^{\rm
local}<14.3$ (95\% C.L.)  \cite{Ade:2013ydc} for reasonable values of $N_0$.

\subsection{Induced bispectrum $\langle\zeta_0\zeta_B\zeta_B\rangle$}\label{BMMzzbzb}

In a similar way, substituting the $C_B$ obtained from (\ref{P_B1})
and (\ref{PBas}) into equation (\ref{fzbb1}) we find that the
non-linearity parameter induced by the cross correlator
$\langle\zeta_0\zeta_B\zeta_B\rangle$ in the squeezed limit $k_1\ll
k_2\sim k_3$ is given by
  \baq
  \label{fzbb}\nonumber
  \fnl^{\zeta\zeta_{B}\zeta_{B}}&=&
  -\frac{10}{6}n{\cal P}_{\zeta}\left(\frac{4^{n+1}\Gamma^2(n+1/2)}{6\pi}\right)^2\times\\\nonumber
  &&
   \left[
  \rule{0pt}{4ex}\right.
  k_3^3\left(C_1(n)+C_2(n)e^{(2n-4)(N_0-N_2)}\right)e^{(2n-4)(N_2-N_3))}
  \left(\frac{(1-e^{(2n-1)N_2})(1-e^{(2n-1)N_3})}{(4-2n)^2}\right)
  \\\nonumber
  &&
  +C_3(n)k_3^3\left(1+e^{(2n-1)(N_3-N_2)}
  \rule{0pt}{3ex}\right.\left)
  \rule{0pt}{3ex}\right.\frac{1}{(2n-1)(4-2n)}\left(
  \rule{0pt}{3ex}\right.
  1-e^{(2n-4)N_2}\left)
  \rule{0pt}{3ex}\right.\\
  &&+C_3(n)
  \frac{2k_3^3}{3(2n-1)}(1+e^{-3 N_{2}})+5 {\rm p.}
  \left]\rule{0pt}{4ex}\right.
  \times
  \left[
  k_1^3+k_2^3+k_3^3
  \right]^{-1}
  \ .
  \eaq

For the special case of a flat spectrum $n=2$ for magnetic fields,
the result takes the form
  \beq
  \label{fzbb_flat}
  \fnl^{\zeta\zeta_{B}\zeta_{B},n=2}=-640\,{\cal
  P}_{\zeta}\left(\frac{k_3^3N_3N_2(N_0-N_2)+5{\rm p.}}{k_1^3+k_2^3+k_3^3}\right)\ .
  \eeq
Using that within the squeezed limit approximation
$k_1\ll k_2\sim k_3$  and using ${\cal
P}_{\zeta}=2.4\times 10^{-9}$ we get the result
  \beq
  \fnl^{\zeta\zeta_{B}\zeta_{B},n=2}\sim -16\times 10^{-7}\,N_2\left(N_2(N_0-N_1)+N_1(N_0-N_2)\right)\ .
  \eeq
which is also within observational bounds for reasonable values of $N_0$, $N_1$ and $N_2$

\subsubsection*{Folded shape}

 Another important shape is the flattened limit $k_1 = 2k_2=2k_3$, where it was earlier found that the magnetic non-linearity parameter, $b_{NL}$, can be large.  It was calculated in \cite{Jain:2012vm} that  $b_{NL}\sim 5760$ in this shape, and the dominating contribution to the cross-correlation function comes from
\bea\label{A&Bfolded}
&& A \approx \hat p\cdot \hat r \frac{k_1^3}{(p r)^{3/2}} 3(n_B-4)\log(-(k_1+p+r) \tau_I)~,\nonumber\\&&  D \approx -\frac{k_1^3}{(p r)^{3/2}} 3(n_B-4)\log(-(k_1+p+r)  \tau_I)~,\qquad J=0,
\eea
as discussed in appendix \ref{appzBB}.

With this ansatz, we can carry out the angular integrals in the flattened limit $k_1 = 2k_2=2k_3$ in the scale invariant case. The angular integrals then gives the leading contribution
 \baq
  \label{zb2b2flat}
   \langle\zeta_0(\k_1){\bf B}^2(\k_2){\bf B}^2(\k_3)\rangle_{\rm c}
  &=&
  20(2\pi)^3\delta(\k_1+\k_2+\k_3)P_{\zeta}(k_1)P_{\B}(k_2)\left(\frac{k_1}{k_3}\right)^3{\mathcal{P}_{\zeta}}\\
  & &\times\frac{3 H^4}{4\pi^2 \lambda } 3(n_B-4)(N_0-N_{\rm CMB})\nonumber
  \eaq
from which we can obtain

\bea
  \label{zzbzbfold}
   \langle\zeta_0(\k_1){\zeta_B}(\k_2){\zeta_B}(\k_3)\rangle
  &=&
  (2\pi)^3\delta(\k_1+\k_2+\k_3)P_{\zeta}(k_1)P_{\zeta}(k_2)\left(\frac{k_1}{k_3}\right)^3\frac{\mathcal{P_{\zeta_B}}}{\mathcal{P}_{\zeta}}\\
  &&\times -\frac{15}{2}(n_B-4)(N_0-N_{\rm CMB}) \nonumber
\eea
If we insert the expression for $\mathcal{P_{\zeta_B}}$, we obtain
 \beq\label{fnlzzbzb}
| \fnl^{flat ~(\zeta\zeta_{B}\zeta_{B})} |\sim 11520 {\cal P}_{\zeta}N_{\rm CMB}^2\,(N_0-N_{\rm CMB})^2
  \ .
  \eeq
Taking $P_{\zeta} = 2.4\times 10^{-9}$ and $N_{CMB} = 60$, we note that $N_0=70$ already induces large non-Gaussianity.

\subsection{Induced bispectrum
$\langle\zeta_B\zeta_B\zeta_B\rangle$}

The integral in (\ref{zbzbzb_int}) can be evaluated using the
magnetic spectrum (\ref{PBas}) which gives the time evolution of the
magnetic fields on superhorizon scales. For $n>1/2$ and for momentum
configurations with $k_i\sim k$, the induced three point function is
given by \cite{Fujita:2013qxa}
  \baq
  \langle\zeta_B(\k_1)\zeta_B(\k_2)\zeta_B(\k_3)\rangle_{\rm c}
  &\simeq&(2\pi)^3\delta(\k_1+\k_2+\k_3)P_{\zeta}(k_1)P_{\zeta}(k_2)
  {\cal P}_{\zeta}\left(\frac{4^{n+1}\Gamma^2(n+1/2)}{3\pi}\right)^3
  \\\nonumber&&\frac{2+2{\rm cos}^2(\k_1,\k_2)}{3(4-2n)^4}\left(1-e^{(4-2n)(N_{\rm CMB}-N_0)}\right)
  \left(e^{-(4-2n)N_{\rm CMB}}-1\right)^3+2{\rm p}.
  \eaq
The corresponding contribution to the non-linearity parameter $\fnl$
reads
  \baq \nonumber
  \fnl^{\zeta_B\zeta_B\zeta_B}&\simeq&{\cal P}_{\zeta}\left(
  \frac{4^{n+1}\Gamma^2(n+1/2)}{3\pi}\right)^3 \frac{5}{9(4-2n)^4}\left(1-e^{(4-2n)(N_{\rm CMB}-N_0)}\right)
  \left(e^{-(4-2n)N_{\rm CMB}}-1\right)^3\times\\
  &&\frac{k_3^3(1+{\rm cos}^2(\k_1,\k_2))+2{\rm
  p.}}{k_1^3+k_2^3+k_3^3}\ .
  \eaq

In the limit of a flat spectrum for magnetic fields $n=2$ and for
the folded shape $\vect k\equiv \vect k_1 = -2 \vect k_2 = -2\vect
k_3$, we then obtain
 \beq
 | \fnl^{\zeta_B\zeta_B\zeta_B, n=2}|=1536{\cal P}_{\zeta}N_{\rm CMB}^3(N_0-N_{\rm CMB})
  \ .
  \eeq
which is shown in the rightmost panel of Figure \ref{fig2} and Figure \ref{fig3} and compared with $ \fnl^{\zeta \zeta_B\zeta_B, n=2}$ in (\ref{fnlzzbzb}) shown in the leftmost panel of Figure \ref{fig2} and Figure \ref{fig3}.
\begin{figure}[!tphb]
\begin{center}
\includegraphics[width=7.5cm]{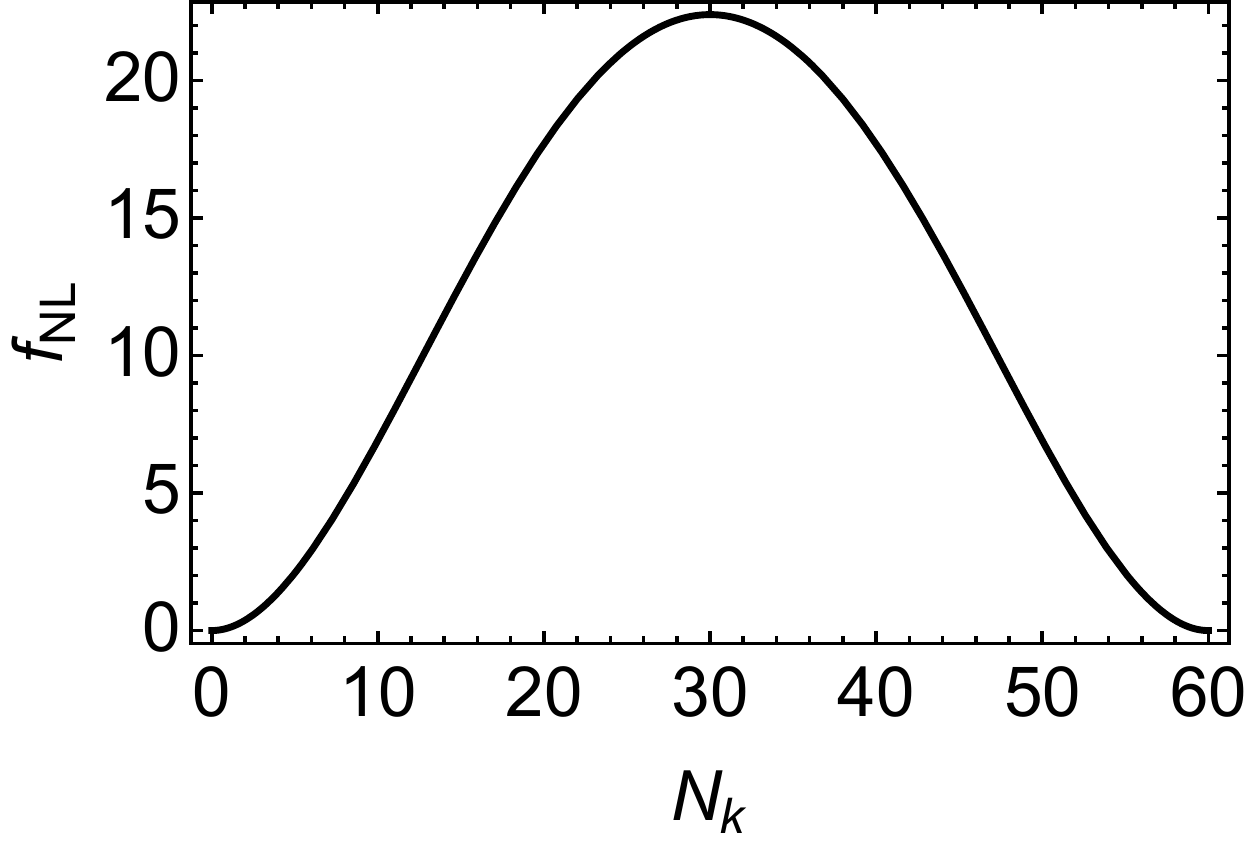}
\includegraphics[width=7.5cm]{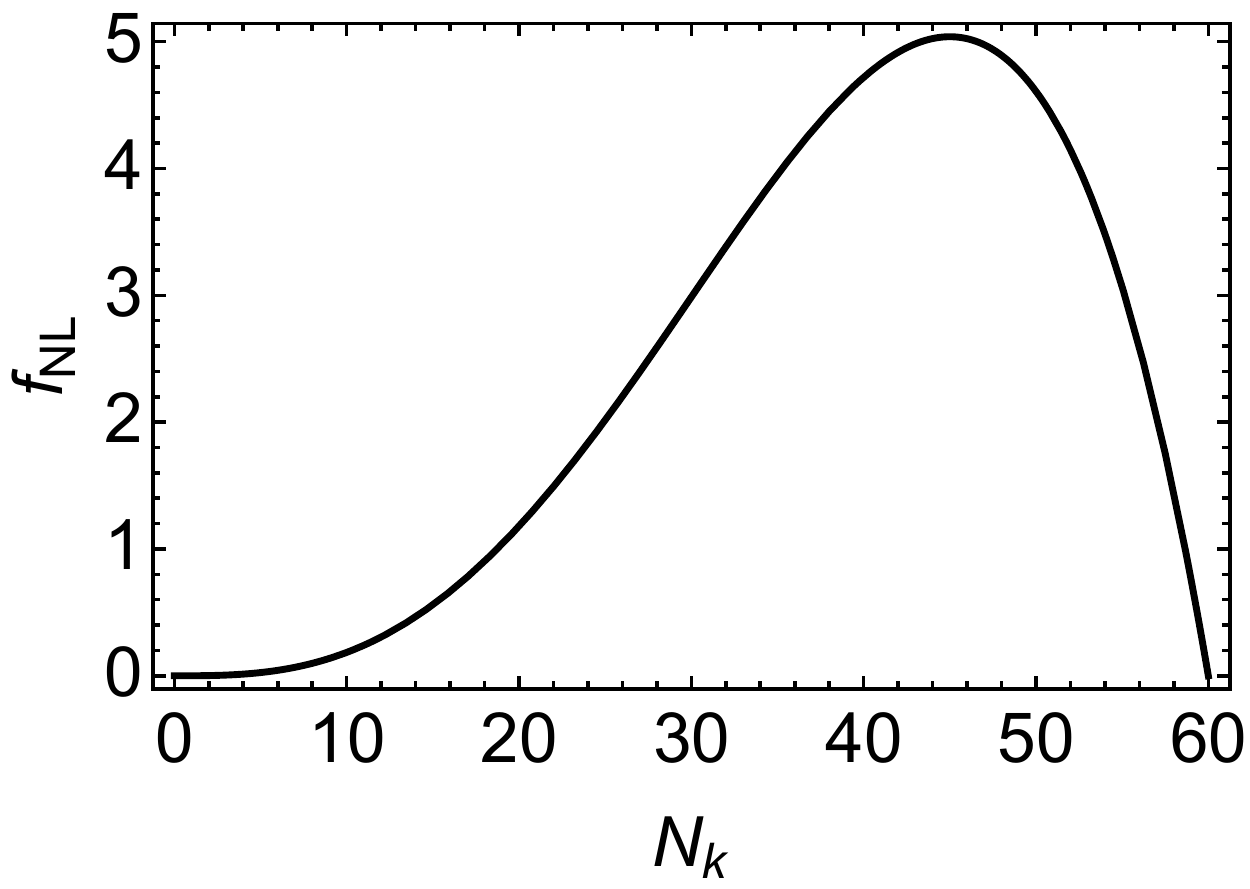}
\end{center}
\caption{The induced $|f_{NL}|$ as a function of the scale $N_k = \ln(k/a_0H_0)$ in the case where the total number of e-folds is $N_{total} = 60$. The left panel is $|f_{NL}^{\zeta \zeta_B\zeta_B}|$ and the right panel is $|f_{NL}^{\zeta_B \zeta_B\zeta_B}|$.}
\label{fig2}
\end{figure}

\begin{figure}[!tphb]
\begin{center}
\includegraphics[width=7.5cm]{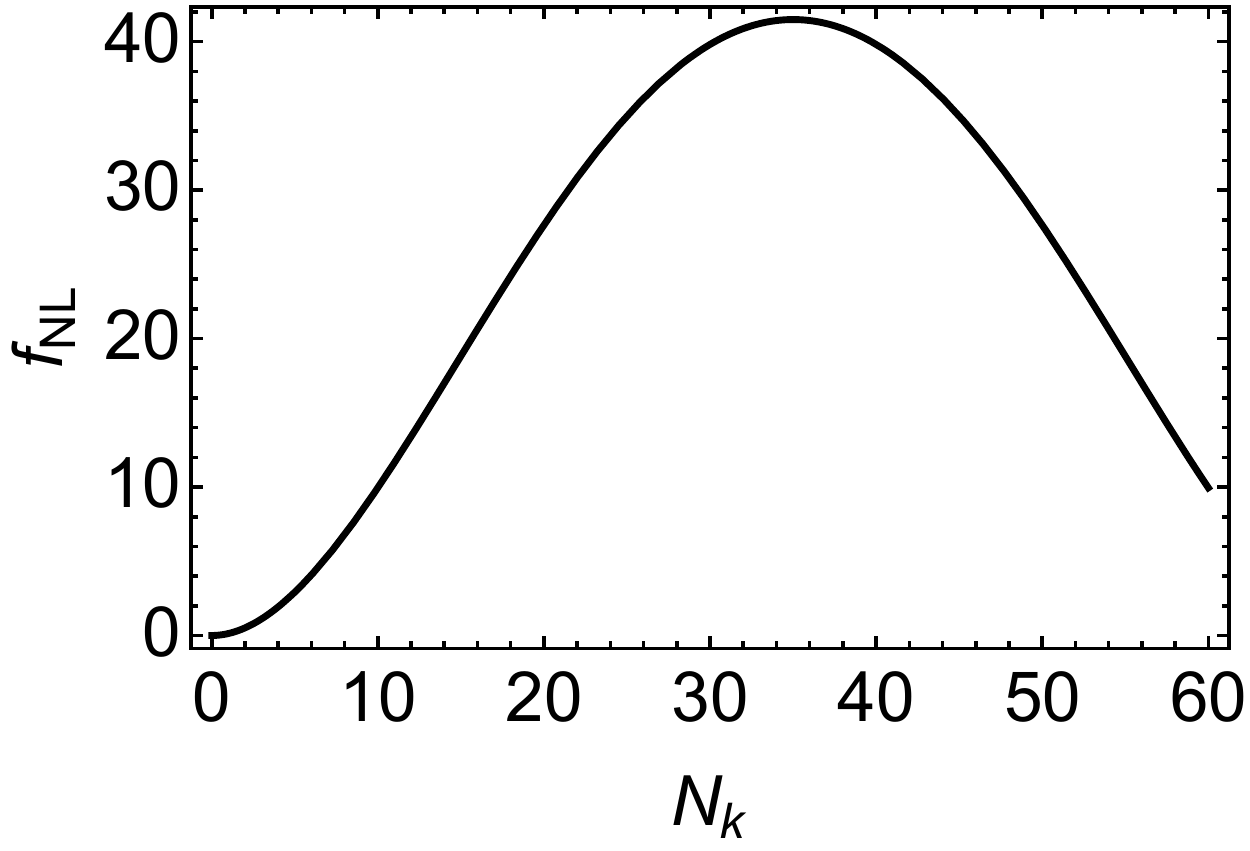}
\includegraphics[width=7.5cm]{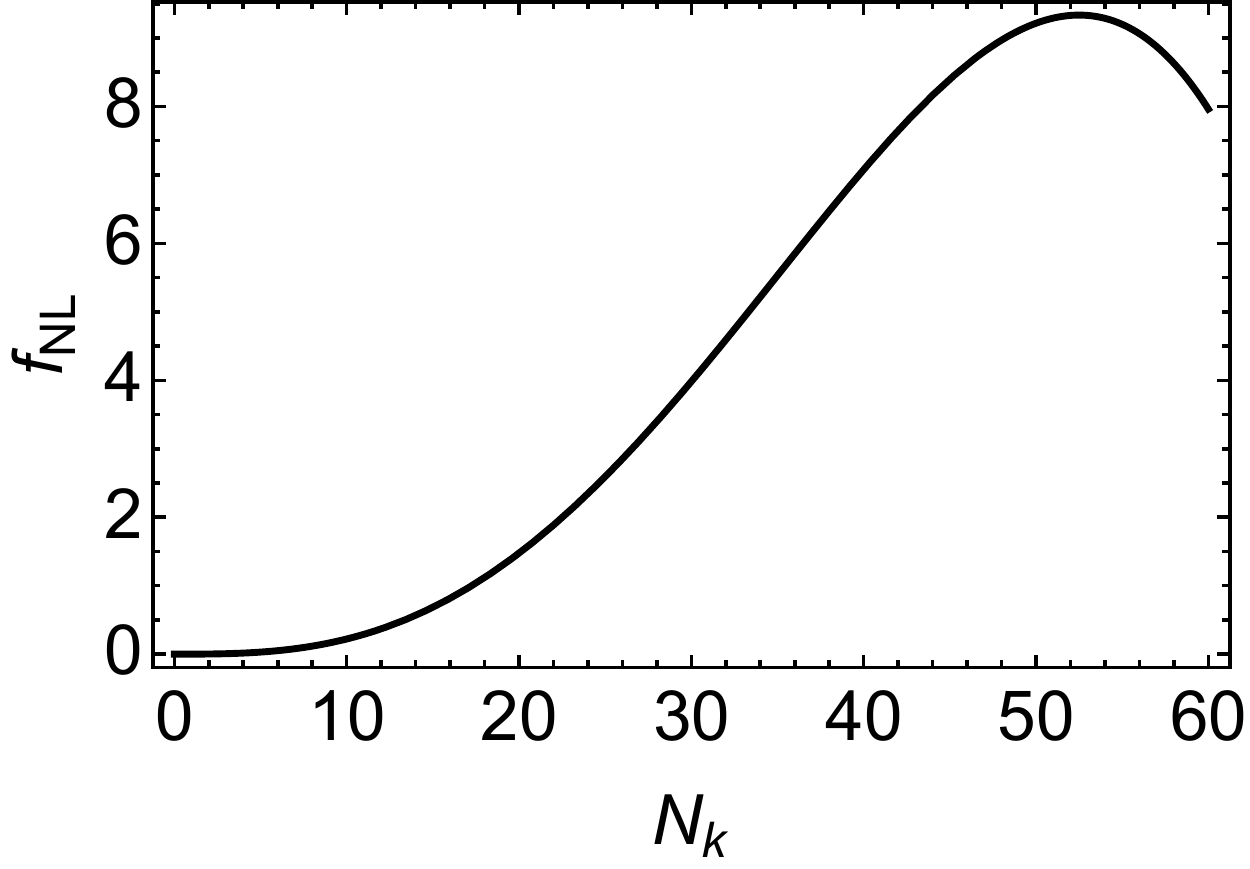}
\end{center}
\caption{The induced $|f_{NL}|$ as a function of the scale $N_k = \ln(k/a_0H_0)$ in the case where the total number of e-folds is $N_{total} = 70$. The left panel is $|f_{NL}^{\zeta \zeta_B\zeta_B}|$ and the right panel is $|f_{NL}^{\zeta_B \zeta_B\zeta_B}|$}
 \label{fig3}
\end{figure}

We see that for very moderate amount of total inflation, slightly more that the required $60$ e-folds, the new non-Gaussian contribution to the CMB from $ \fnl^{\zeta \zeta_B\zeta_B, n=2}$ in (\ref{fnlzzbzb}) can be very large on CMB scales, and potentially provide very strong constraints on the model.

\section{Summary and conclusions}

We have studied the constraints on gauge field production during inflation imposed by requiring that their effect on the CMB anisotropies are subdominant.
Focussing on the non-Gaussianity induced by the gauge field production, we studied for the first time the bispectrum of the primordial curvature perturbation induced by the cross correlation between the curvature perturbation induced by the inflaton and the curvature perturbation induced by the magnetic field, defined by the gauge field. In order to make this study as model-independent as possible, we used a general parametrization of the cross correlation between the magnetic field, and the primordial  curvature perturbation in terms of the magnetic non-linearity parameters. In order to facilitate this parametrization, we have defined the magnetic non-linearity parameters $\be_{NL}$, $c_{NL}$ characterizing the strength of the four point function $\left<\zeta_0\zeta_0 B^2\right>$, in addition to the non-linearity parameter $b_{NL}$ parametrizing the strength of the three-point cross-correlation function $\left<\zeta_0 B^2\right>$. In appendix \ref{appPB} the non-linearity p!
 arameters were computed in the squeezed limit.

Since the magnetic field squared $B^2$ acts as a non-Gaussian
iso-curvature perturbation during inflation, it induces a
non-Gaussian primordial curvature perturbation $\zeta_B$. As a
measure of this non-Gaussianity, we have computed the induced
primordial curvature bispectrum from the contributions of the form
$\left<\zeta_0 \zeta_0 \zeta_B\right>$, $\left<\zeta_0 \zeta_B
\zeta_B\right>$ and $\left<\zeta_B \zeta_B \zeta_B\right>$. The
first two of these depend on $b_{NL}$ and $c_{NL}$ and can be used
to derive observational constrains on the magnetic non-linearity
parameters. Assuming a power law parametrization for the spectrum of
the magnetic fields produced during inflation but treating the
coupling $\lambda(\phi)$ as a free function, we have then derived
the observational constraints on $b_{NL}$ and $c_{NL}$.

In particular we have shown that in a general class of models, the
new contribution to the bispectrum of the primordial curvature
perturbation from $\left<\zeta_0 \zeta_B \zeta_B\right>$  can be the
dominant source of non-Gaussianity and lead to a large non-Gaussian
contribution in the folded shape if inflation last only slightly
longer than the required $60$ e-folds. This implies new strong
phenomenological constraints on gauge field production in this class
of models when compared with the absence of a non-Gaussian
primordial signal as observed by the Planck satellite
\cite{Ade:2013ydc}.

If inflation last much longer than the observable $60$ e-folds, the
results presented here will provide the average correlation function
in the full inflated volume, while the observed correlation function
may deviate from this value
\cite{Giddings:2010nc,Byrnes:2010yc,Giddings:2011zd,Byrnes:2011ri,Nurmi:2013xv,Byrnes:2013qjy,Nelson:2012sb,LoVerde:2013xka,LoVerde:2013dgp,Bramante:2013moa}.
In this case, one should treat the long wavelength modes as a
homogenous background  for the shorter wavelength modes within the
observable region, which by its vector nature breaks isotropy. In
that case effects similar to those discussed here leads to further
new constraints on the magnetic non-linearity parameters by their
anisotopic contribution to the power spectrum and bispectrum of the
curvature perturbation. The analogous $b_{NL}$ independent
effects was discussed in \cite{Bartolo:2012sd} (see also
\cite{Abolhasani:2013zya,
Shiraishi:2013vja,Emami:2013bk,Soda:2012zm}). While it is beyond the
scope of the present work, it would be interesting in the future to
study also the new sources of anisotropy from the cross correlation
functions of the magnetic field with the inflaton.
\newline
\newline
{\bf Acknowledgments:} SN is supported by the Academy of Finland
grant 257532 and MSS is support by a Lundbeck Foundation Jr. Group Leader Fellowship.

\appendix

\section{Source term in the scale-invariant limit}\label{source}

From the definition of the curvature perturbation in terms of the inflaton and gauge field curvature perturbations
\beq\label{zetadef}
\zeta = \zeta_\phi + \tilde\zeta_B
\eeq
we have the equations governing their time-evolution derived in (\ref{dotzphi}) in the introduction
\beq\label{dotzeta}
\dot\zeta_\phi = H\frac{Q}{\dot\rho}~, \qquad \dot{\tilde\zeta}_B = -H\frac{Q}{\dot\rho} -\frac{H}{\rho+p}\delta P_{nad}
\eeq
where
\beq \label{Pnad}
\delta P_{nad} = \delta p_B-\frac{\dot p}{\dot \rho}\delta\rho_B =  \frac{4}{3}\delta\rho_B~.
\eeq

It follows from (\ref{zetadef}) and (\ref{dotzeta}) that only if
\beq \label{cond1}
H\frac{Q}{\dot\rho}= -\frac{H}{\rho+p}\delta P_{nad}~,
\eeq
it is consistent to assume
\beq
\dot\zeta =  \dot\zeta_\phi + \dot{\tilde\zeta}_B\approx \dot\zeta_\phi = H\frac{Q}{\dot\rho}~,
\eeq
like in \cite{Barnaby:2012tk, Lyth:2013sha}, instead of using the more generally valid expression
\beq
\dot\zeta = -\frac{H}{\rho+p}\delta P_{nad}~.
\eeq

Since the source term, $Q$, is given by
\beq
Q =\frac{\dot \lambda}{\lambda}\delta\rho_B
\eeq
then with the assumption of a power-law behavior $\lambda =\lambda_0 (a/a_0)^{2n}$, such that $\dot\lambda/\lambda =2n H$, we have that the condition (\ref{cond1}) for the approximations in \cite{Barnaby:2012tk, Lyth:2013sha} to be valid, becomes equivalent the the condition
\beq
2n \frac{H^2}{\dot\rho}\delta\rho_B=- \frac{4}{3}\frac{H}{\rho+p}\delta\rho_B
\eeq
where we used (\ref{Pnad}). Now using that $\dot\rho =-3(\rho+p)$, we have that this condition is only satisfied in the flat case when $n=2$. This explains why \cite{Barnaby:2012tk, Lyth:2013sha} finds the right spectrum $P_{\zeta_B}$ in the flat limit, even if their treatment is generally formally inconsistent.

\section{Parametrization of $P_B(k)$ and local magnetic non-linearity parameters}\label{appPB}

We briefly review the magnetic consistency relation for $b_{NL}^{local}$  \cite{Jain:2012ga}, and generalize it to $c_{NL}^{local}$.

Let us consider the basic correlation function $\left<\zeta_0(\tau_I,\vect k_1) A_i(\tau_I,\vect k_2) A_j(\tau_I,\vect k_3)\right>$ in the squeezed limit $k_1 \ll k_2,k_3$. In this limit, the only effect of  the long wavelength mode $\zeta_0(\tau,\vect k_1)$ is to locally rescale the background as $a\to a_B=e^{\zeta_B}a$ when computing the correlation functions on shorter scales given by $k_2, k_3$, and one can therefore as usual write
\bea\label{rel1}
&&\lim_{k_1,\dots,k_n \to 0} \left< \zeta_0(\tau_I,\vect k_1)\dots \zeta_0(\tau_I,\vect k_n)A_i(\tau_I,\vect k_{n+1}) A_j(\tau_I,\vect k_{n+2})\right> \nonumber\\
&&\qquad= \left< \zeta_0(\tau_I,\vect k_1)\dots \zeta_0(\tau_I,\vect k_n)\left< A_i(\tau_I,\vect k_{n+1}) A_j(\tau_I,\vect k_{n+2})\right>_{\zeta_0}\right>.
\eea
Here $\left<A_i(\tau_I,\vect k_2) A_j(\tau_I,\vect k_3)\right>_{\zeta_0}$ is the correlation function of the short wavelength modes in the background of the long wavelength modes of $\zeta_0$ .

Since the equations of motion of the gauge field are conformal invariant in the absence of the coupling $\lambda(\phi)$, it follows that the gauge field only feels the background expansion through the coupling $\lambda$, where  $\lambda$ depends on the scale factor through $\phi$. Using that the gauge field scales like $1/\sqrt\lambda$, then in order to evaluate the correlation function for a non-trivial $\lambda$, one can then write $A_i(\tau,\vect k)$ in terms of the Gaussian field $A_i(\tau,\vect k)=\sqrt{(\lambda_0/\lambda)}A_i^{(G)}(\tau,\vect k)$, where the Gaussian gauge field $A_i^{(G)}(\tau,\vect k)$ is defined with a homogeneous background coupling, $\lambda_0$,  and then expand $\lambda = \lambda(a)$ around the homogenous background value,
\beq\label{dN}
\lambda = \lambda_0 + \frac{d \lambda_0}{d\ln a}\delta\ln a + \frac{1}{2}\frac{d^2 \lambda_0}{d\ln a^2}\delta\ln a^2 +\dots = \lambda_0+\frac{d \lambda_0}{d\ln a} \zeta_0+ \frac{1}{2}\frac{d^2 \lambda_0}{d\ln a^2}\zeta_0^2+\dots ~.
\eeq
which yields
\beq\label{AB}
A_i(\tau,\vect k)= A_i^{(G)}(\tau,\vect k)\left(1-\frac{1}{2}\frac{1}{H}\frac{\dot\lambda}{\lambda}\zeta_0+\left(\frac{3}{8}\frac{1}{H^2}\frac{\dot\lambda^2}{\lambda^2}-\frac{1}{4}\frac{1}{H^2}\frac{\ddot\lambda}{\lambda}\right)\zeta_0^2+\dots\right)
\eeq
By comparison with the definitions of $b_{NL}$ and $c_{NL}$ in
equation (\ref{Blocal}), we conclude that
\beq
\label{bnl_exp}
b_{NL}= -\frac{1}{H}\frac{\dot\lambda}{\lambda}
\eeq
and
\beq
\label{cnl_exp}
c_{NL}=\frac{9}{4}\frac{1}{H^2}\frac{\dot\lambda^2}{\lambda^2}-\frac{3}{2}\frac{1}{H^2}\frac{\ddot\lambda}{\lambda}~.
\eeq

Finally by inserting the expansion (\ref{AB}) into (\ref{rel1}), we can reproduce the consistency relations
\bea \label{zBB2}
& & \lim_{k_1\to 0} \left<\zeta_0(\tau_I,\vect{k}_1){\bf B}(\tau_I,\vect{k}_2)\cdot {\bf B}(\tau_I,\vect{k}_3)\right> \nonumber \\
& &\qquad= -\frac{1}{H}\frac{\dot\lambda}{\lambda}(2\pi)^3\delta^{(3)}(\vect{k}_1+\vect{k}_2+\vect{k}_3)P_\zeta(k_1) P_B(k_2)
\eea
and
\bea \label{zzBB2}
& & \lim_{k_1, k_2\to 0} \left<\zeta(\tau_I,\vect{k}_1)\zeta_0(\tau_I,\vect{k}_2){\bf B}(\tau_I,\vect{k}_3)\cdot {\bf B}(\tau_I,\vect{k}_4)\right> \nonumber \\
& &\qquad = \left(2\frac{1}{H^2}\frac{\dot\lambda^2}{\lambda^2}-\frac{1}{H^2}\frac{\ddot\lambda}{\lambda}\right)(2\pi)^3\delta^{(3)}(\vect{k}_1+\vect{k}_2+\vect{k}_3+\vect{k}_4)P_\zeta(k_1)P_\zeta(k_2) P_B(k_3)
\eea

Using the power-law assumption for the coupling in terms of the
scale-factor, $\lambda \propto a^{2n}(t)$, when then obtain from (\ref{bnl_exp}) and (\ref{cnl_exp})
\beq
b_{NL} = -2 n = n_B-4~, \qquad c_{NL} =\frac{3}{4} b_{NL}^2\ .
\eeq
Therefore, in this case the magnetic non-linearity parameters are
fully determined by the exponent. The magnetic spectrum also takes a
power law form $P_{B}\propto k^{1-2n}$ as can be seen in its
explicit expression (\ref{PBas}).

In most of the paper we have for simplicity assumed the magnetic spectrum, $P_B(k)$, to have a power law form. If this is derived from assuming that the coupling, $\lambda$, also have the simple power law form, then we have just seen how $b_{NL}$ and $c_{NL}$ are fixed the power law index $n$.  If the power law assumption for $\lambda$ was the only way to obtain a power law form for the magnetic spectrum, $P_B(k)$, it would therefore not make much sense to constrain $b_{NL}$ and $c_{NL}$ in a model independent way. Here we will therefore demonstrate with a concrete example that a power law
spectrum $P_{B}$ (\ref{P_B1}) can be generated also for more general
couplings $\lambda(a)$ which are not of the simple power law form.
In this case the magnetic non-linearity parameters $\bnl$ and $\cnl$
are in general independent of each other and their magnitudes are
not determined by the properties of the spectrum.

We work in the Coulomb gauge specified by the conditions $A_0=0$ and
$\partial_i A_i=0$. The spatial part of the vector potential is then
decomposed in the standard way
  \beq
  A^i_{\k}=\sum_{\sigma=\pm}\left(
  \epsilon_{\sigma}^i(\k)e^{i\k\cdot\x}\hat{a}_{\sigma}(\k)A_{\k}+h.c.\right)\
  ,
  \eeq
where the polarization operators satisfy $k_i\epsilon_i^{\sigma}=0,~
\epsilon_i^{\sigma}{\epsilon_i^{\sigma'}}^{*}=\delta^{\sigma\sigma'},~\epsilon_i^{\sigma}{\epsilon_j^{\sigma}}^{*}=
\delta_{ij}-k_ik_j/k^2$. The commutation relations of the creation
/annihilation operators are given by
$[\hat{a}^{\sigma}_{\k},\hat{a}^{\sigma'}_{\k'}{}^{\dag}]=(2\pi)^2\delta(\k-\k')\delta^{\sigma\sigma'}$.
The equation of motion is then given by
  \beq
  \label{Aeom}
  (\sqrt{\lambda}A_{\k})''+\left(k^2-\frac{(\sqrt{\lambda})''}{\sqrt{\lambda}}\right)\sqrt{\lambda}A_{\k}=0\
  ,
  \eeq
where the prime denotes a derivative with respect to the conformal
time $\eta=-1/(aH)$. In the superhorizon limit $k\ll -1/\eta$ the
gradient term in (\ref{Aeom}) can be neglected and the equation of
motion recast in the simple form
  \beq
  \label{Aeom_superhor}
  (\lambda A'_{\k})'=0\ .
  \eeq

Using the definitions above we can expand the coupling $\lambda(a)$
as
  \beq
  \label{lambda_exp}
  \lambda(a)=\lambda_0\left(1-\bnl{\rm ln}\frac{a}{a_0}+\left(\frac{3}{4}\bnl^2-\frac{1}{3}\cnl\right)
  {\rm ln}^2\frac{a}{a_0}+\ldots\right)\ .
  \eeq
Substituting this into (\ref{Aeom_superhor}) we can express the
superhorizon solution for the vector potential in the integral form
  \beq
  A_{\k}(a)=\frac{D}{a_0H\lambda_0}\int\limits_{0}^{{\rm
  ln}(a/a_0)}\frac{\d x}{e^{x}(1-\bnl
  x-(3\bnl^2/4-\cnl/3)x^2+\ldots)}\ .
  \eeq
where $D$ is a constant to be determined by matching with the
subhorizon solution. Setting now $\cnl = 9 \bnl^2/4$ so that the
second order term in (\ref{lambda_exp}) vanishes, neglecting the
higher order corrections denoted by the ellipsis, we obtain the
result
  \beq
  A_{\k}(a)=
  \frac{D}{a_0H\lambda_0}\frac{e^{1/\bnl}}{\bnl}\left({\rm Ei}(-1/\bnl-{\rm
  ln}(a/a_0))-{\rm Ei}(-1/\bnl)\right)\ .
  \eeq
Here ${\rm Ei}(z)$ denotes the exponential integral ${\rm
Ei}(z)=-\int_{-z}^{\infty}\d t e^{-t}/t$. To determine the constant
$D$ we match this superhorizon solution with the subhorizon
solution $A_{\k}^{\rm sub}=e^{ik\eta}/\sqrt{k\lambda(\eta)}$ at
horizon crossing by setting $|A_{\k}(a=k/H)|=|A_{\k}^{\rm
sub}(a=k/H)|$. This renders the superhorizon result in the form
  \beq
  A_{\k}(a\gg k/H) =\frac{1}{\sqrt{k\lambda(k/H)}}\frac{{\rm Ei}(-1/\bnl-{\rm
  ln}(a/a_0))-{\rm Ei}(-1/\bnl)}{{\rm Ei}(-1/\bnl-{\rm
  ln}(k/k_0))-{\rm Ei}(-1/\bnl)}\approx \frac{1}{\sqrt{k\lambda(k/H)}}\ ,
 \eeq
where the last approximative form holds in the limit $k\gg k_0$ and
$k_0 = a_0 H$. Taking the limit $k\gg k_0$ of the superhorizon
result implies that $a/a_0\gg k/(a_0H)\gg 1$. In other words this
corresponds considering modes $k$ bigger than the expansion scale
$k_0$ long after the horizon crossing of the mode $k_0$.

Using the asymptotic result for the vector potential
$A_{\k}(a)\approx 1/\sqrt{k\lambda(k/H)}$ we can then work out the
corresponding spectrum of magnetic fields. The result is given by
  \baq
  P_{B}(\eta,k)&=&2\frac{k^2}{a^4}|A_{\k}(\eta)|^2\\\nonumber
  &=&\frac{2H^4}{\lambda(\eta)}\frac{1-\bnl{\rm ln}(-k_0\eta)}{1-\bnl{\rm
  ln}(k/k_0)}(-\k\eta)^4 k^{-3}\\\nonumber
  &\approx&\frac{2H^4}{\lambda(\eta)}(-\k\eta)^{4-1/{\rm
  ln}(k/k_0)} k^{-3}\ ,
  \eaq
where in the last step we have used that ${\rm ln}(k/k_0)\gg 1$.
Therefore, in this limit we find that the power spectrum of the
magnetic fields is approximatively of the power law form (\ref{P_B1})
even if the coupling is given by $\lambda=\lambda_0(1-\bnl {\rm
ln}(a/a_0))$ instead of a power law $\lambda\propto a^n$.

As a an explicit toy example it shows, that if we choose the coupling constant to be a constant up to a logarithmic correction, then as expected it reproduce the spectrum with a constant coupling function up to logarithmic corrections, but interestingly now $b_{NL}$ is a free parameter. Although the generated magnetic field at the end of inflation is small in this model, and therefore not of great phenomenological interest, it serves as a useful demonstration model for the purpose of showing that in general $\bnl$ and $\cnl$ should treated as free parameters in a general treatment.

In fact, as also discussed in the introduction, we might expect that the relation between the form of the coupling, $\lambda$, and the magnetic non-linearity parameters, will be different in models with deviations from the Bunchs-Davis vacuum or with extra degrees of freedom.

\section{The tensor structure of the cross-correlation bispectrum}\label{appzBB}

Analogous to (\ref{zBB0}), it is convenient to introduce also a tensor bispectrum, where the magnetic fields are left uncontracted
\beq \label{zBBt0}
\left<\zeta(\vect{k}_1)B_i(\vect{k}_2)  B_j(\vect{k}_3)\right> \equiv(2\pi)^3\delta^{(3)}(\vect{k}_1+\vect{k}_2+\vect{k}_3)B_{ij}^{\zeta B B}(\vect{k}_1,\vect{k}_2, \vect{k}_3) ~.
\eeq

The tensor cross-correlation bispectrum of the curvature perturbation with the magnetic field, is constructed from the more fundamental correlation function of the curvature perturbation with the vector field itself $\left<\zeta(k_1)A_i(k_2)A_j(k_3)\right>$, which places some constraints on its general form. We will assume that  $\left<\zeta(k_1)A_i(k_2)A_j(k_3)\right>$ is a tensor function of $\hat k_2$ and $\hat k_3$
\bea \label{zBBt1}
\left<\zeta(\vect{k}_1)A_i(\vect{k}_2)  A_j(\vect{k}_3)\right> &=& (2\pi)^3\delta^{(3)}(\vect{k}_1+\vect{k}_2+\vect{k}_3) \left[ A\delta_{ij} + B(\hat k_{2i}\hat k_{2j}+\hat k_{3i}\hat k_{3j})\right.\nonumber\\
& & +C\hat k_{2i}\hat k_{3j}+D\hat k_{2j}\hat k_{3i} +E \hat k_{2i} (\hat k_{2}\times\hat k_{3})_j+ F\hat k_{2j} (\hat k_{2}\times\hat k_{3})_i\nonumber\\
& &+G \hat k_{3i} (\hat k_{2}\times\hat k_{3})_j+ H\hat k_{3j} (\hat k_{2}\times\hat k_{3})_i\nonumber\\
& &\left.+J (\hat k_{2}\times\hat k_{3})_i (\hat k_{2}\times\hat
k_{3})_j)\right]|\zeta_{k_1}|^2|A_{k_2}||A_{k_3}|~. \eea where
$\zeta_{k}$ and $A_k$ are the mode functions of the curvature
perturbation and the vector field respectively. Using that the
correlation function is invariant under the exchange of $A_i(k_2)$
and $A_j(k_3)$, we have $E=F$ and $G=H$, and using
\beq \label{zBBt2}
\left<\zeta(\vect{k}_1)B_i(\vect{k}_2)  B_j(\vect{k}_3)\right>= \ep_{ilk}\ep_{jmn}k_2^lk_3^m\left<\zeta(\vect{k}_1)A^k(\vect{k}_2)  A^n(\vect{k}_3)\right>~,
\eeq
we obtain
\bea
B_{ij}^{\zeta B B} &=& \left[ A(\delta_{ij}\delta_{lm}-\delta_{im}\delta_{lj})\hat k_{2l}\hat k_{3m}+D(\hat k_2\times \hat k_3)_i(\hat k_2\times \hat k_3)_j\right.\nonumber\\& &+G((\hat k_2\times \hat k_3)_i(\hat k_{2 } - \hat k_{3} \hat k_2\cdot \hat k_3)_j-(\hat k_2\times \hat k_3)_j(\hat k_{2 } - \hat k_{3} \hat k_2\cdot \hat k_3)_i)\nonumber\\
& &\left.+J(\hat k_2 \hat k_2\cdot\hat k_3-\hat k_3)_i(\hat k_2-\hat k_3 \hat k_2\cdot\hat k_3)_j\right]P_{\zeta}(k_1)\sqrt{P_B(k_2)P_B(k_3)}~.
\eea
The trace of the magnetic non-linearity parameter $b_{NL}$ is given by the trace of $B_{ij}^{\zeta B B} $,
\beq
b_{NL}= 2\frac{\textrm{Tr}(B^{\zeta BB})}{P_{\zeta}(k_1)(P_B(k_2)+P_B(k_3))}~,
\eeq
where with $\vect{k}_2\cdot\vect{k}_3=k_2 k_3\cos \theta$, we have
\beq
\textrm{Tr}(B^{\zeta BB}) = (2 A\cos \theta +D \sin^2\theta +J\sin^2\theta\cos\theta)P_{\zeta}(k_1)\sqrt{P_B(k_2)P_B(k_3)}~.
\eeq
In the squeezed limit, we have $\cos\theta=-1$ and in the flattened shape, we have $\cos=1$ and $k_2=k_3$. Thus in these shapes, we have
\beq
b_{NL}^{local} = -2A~,\qquad b_{NL}^{flat} = 2A~.
\eeq
Another simple shape is the orthogonal shape $\cos\theta =0$, for which we have
\beq\label{bnlorth}
b_{NL}^{orthogonal} = 2D \frac{\sqrt{P_B(k_2)P_B(k_3)}}{P_B(k_2)+P_B(k_3)}~.
\eeq
On the other hand the equilateral shape contains contributions from both $A,D$ and $J$.

By noticing the fact that the de Sitter isometries becomes the conformal group on the future boundary of de Sitter space, it has been argued that one can use this conformal symmetry to constrain the asymptotic super horizon structure of the correlation function in $\langle\zeta(\vect{k}_1) A_i(\vect{k}_2) A_j(\vect{k}_3) \rangle $ in (\ref{zBBt1}) \cite{Biagetti:2013qqa}. The result of \cite{Biagetti:2013qqa} obtained with $n=2$, can be reproduced in the current parametrization in (\ref{zBBt1}) by taking $A= -(\hat k_2\cdot \hat k_3)D$ and $B=C=F=G=H=J=0$. This means that in this case by symmetries alone, we can determine that the leading logarithmical divergent contribution at late time, is given by $A$, $D$ up to an overall numerical factor. The precise calculation of the full correlation function shows that in this case, the dominant term in the limit $\log(-k_t\tau) \to \infty$ are
\beq
A \approx \hat k_2\cdot \hat k_3 \frac{k_1^3}{(k_2 k_3)^{3/2}} 3(n_B-4)\log(-k_t \tau_I)~,\qquad  D \approx -\frac{k_1^3}{(k_2 k_3)^{3/2}} 3(n_B-4)\log(-k_t  \tau_I)~,
\eeq
where $k_t=k_1+k_2+k_3$.  One important subtlety of this argument is however that these leading logarithmic terms are suppressed by a factor of $k_1^3$, which vanishes in the exactly squeezed limit $k_1\to 0$. Instead, as mention above, in the squeezeed limit we can identify $A$ with $-b_{NL}^{local}/2$, which can be obtained from the squeezed limit magnetic consistency relation \cite{Jain:2012ga,Jain:2012vm}.

\bibliographystyle{JHEP}
\bibliography{magneticNG}

\end{document}